\documentclass[aps,prd,floats,nofootinbib,showpacs]{revtex4}

\usepackage{latexsym}
\usepackage{amssymb}
\usepackage[dvips]{graphicx}
\usepackage[applemac]{inputenc}
\usepackage[nottoc]{tocbibind} %prevents the entry "Table of Contents" in the "Table of Contents"
\usepackage{float}
\usepackage{fancyhdr}
\usepackage{amsfonts}
\usepackage{amsmath}
\usepackage{color}
\usepackage{mathrsfs}
\usepackage{bm}% bold math
\usepackage{epsfig}% bold math

\newcommand{\be}{\begin{equation}}
\newcommand{\ee}{\end{equation}}

\def\lsim{\mathrel{\raise.3ex\hbox{$<$\kern-.75em\lower1ex\hbox{$\sim$}}}} 
\def\gsim{\mathrel{\raise.3ex\hbox{$>$\kern-.75em\lower1ex\hbox{$\sim$}}}}

\def\beq{\begin{eqnarray}}
\def\eeq{\end{eqnarray}}
\def\bea{\begin{eqnarray}}
\def\eea{\end{eqnarray}}
\def\bit{\begin{itemize}}
\def\eit{\end{itemize}}

\def\tev{\, {\rm TeV}}
\def\gev{\, {\rm GeV}}
\def\mev{\, {\rm MeV}}

\def\cta{\cos\theta_A}

\def\mueff{\mu_\mathrm{eff}}

\def\cta{\cos\theta_A}

\def\tauptaum{\tau^+\tau^-}

\def\sig{\sigma}

\def\ls#1{\ifmath{_{\lower1.5pt\hbox{$\scriptstyle #1$}}}}
\def\lss#1{\ifmath{^{\,\lower2.5pt\hbox{$\scriptstyle #1$}}}}

\def\sig{\sigma}

\def\mz{m_Z}

\def\mhi{m_{\hi}}

\def\h{h}

\def\lam{\lambda}
\def\kap{\kappa}
\def\alam{A_\lam}
\def\akap{A_\kap}

\def\wtil{\widetilde}

\def\tauptaum{\tau^+\tau^-}

\def\lsim{\mathrel{\raise.3ex\hbox{$<$\kern-.75em\lower1ex\hbox{$\sim$}}}}
\def\gsim{\mathrel{\raise.3ex\hbox{$>$\kern-.75em\lower1ex\hbox{$\sim$}}}}
\def\ifmath#1{\relax\ifmmode #1\else $#1$\fi}

\def\third{\ifmath{{\textstyle{1 \over 3}}}}

\def\vev#1{\langle #1 \rangle}
\def\lam{\lambda}

\def\mhi{m_{h_1^0}}

\def\msusy{m_{\rm SUSY}}

\def\hsm{h_{\rm SM}}

\def\hl{h^0}
\def\hh{H^0}
\def\ha{A^0}
\def\hp{h^+}

\def\mhl{m_{\hl}}
\def\mhh{m_{\hh}}
\def\mha{m_{\ha}}
\def\mhp{m_{\hp}}

\def\tanb{\tan\beta}

\def\mz{m_Z}

\def\cnone{\wt\chi^0_1}

\def\cntwo{\wt\chi^0_2}

\def\mcnone{m_{\cnone}}
\def\mcntwo{m_{\cntwo}}

\def\wt{\widetilde}

\def\cpmone{\wt \chi^{\pm}_1}

\def\mcpmone{m_{\cpmone}}

\def\meff{m_{eff}}

\def\MPL #1 #2 #3 {{\sl Mod.~Phys.~Lett.}~{\bf#1} (#3) #2}
\def\NPB #1 #2 #3 {{\sl Nucl.~Phys.}~{\bf #1} (#3) #2}
\def\PLB #1 #2 #3 {{\sl Phys.~Lett.}~{\bf #1} (#3) #2}
\def\PR #1 #2 #3 {{\sl Phys.~Rep.}~{\bf#1} (#3) #2}
\def\PRD #1 #2 #3 {{\sl Phys.~Rev.}~{\bf #1} (#3) #2}
\def\PRL #1 #2 #3 {{\sl Phys.~Rev.~Lett.}~{\bf#1} (#3) #2}
\def\RMP #1 #2 #3 {{\sl Rev.~Mod.~Phys.}~{\bf#1} (#3) #2}
\def\ZPC #1 #2 #3 {{\sl Z.~Phys.}~{\bf #1} (#3) #2}
\def\IJMP #1 #2 #3 {{\sl Int.~J.~Mod.~Phys.}~{\bf#1} (#3) #2}
\def\NIM #1 #2 #3 {{\sl Nucl.~Inst.~and~Meth.}~{\bf#1} {#3} #2}
\def\lam{\lambda}
\def\br{B}
\def\tauptaum{\tau^+\tau^-}

\def\gam{\gamma}

\def\anti{\overline}
\def\epem{e^+e^-}
\def\mupmum{\mu^+\mu^-}

\def\mupmum{\mu^+\mu^-}

\def\ie{{\it i.e.}}

\def\anti{\overline}

\def\ai{a_1}
\def\aii{a_2}
\def\mai{m_{\ai}}
\def\maii{m_{\aii}}

\def\pb{~{\rm pb}}
\def\mev{~{\rm MeV}}
\def\gev{~{\rm GeV}}
\def\tev{~{\rm TeV}}

\def\hi{\h_1}
\def\hii{\h_2}
\def\hiii{\h_3}
\def\mhi{m_{\hi}}
\def\mhii{m_{\hii}}
\def\mhiii{m_{\hiii}}

\newcommand{\nc}{\newcommand}

\nc{\baa}{\begin{array}}      \nc{\eaa}{\end{array}}

\nc{\bce}{\begin{center}}     \nc{\ece}{\end{center}}
\def\beqa{\begin{eqnarray}}
\def\eeqa{\end{eqnarray}}

\def\bmini{\begin{minipage}}
\def\emini{\end{minipage}}
\newcommand{\ba}{\begin{array}}
\newcommand{\ea}{\end{array}}

\def\ee{e^+e^-}

\def\tanb{\tan\beta}

\def\ie{{\it i.e.}}

\def\gam{\gamma}

\def\lam{\lambda}
\def\sig{\sigma}

\def\anti{\overline}

\def\vev#1{{\langle #1 \rangle}}

\def\caibb{C_{\ai b\anti b}}
\def\caiibb{C_{\aii b\anti b}}
\def\chibb{C_{\hi b\anti b}}
\def\chiibb{C_{\hii b\anti b}}
\def\chiiibb{C_{\hiii b\anti b}}

\def\cta{\cos\theta_A}

\def\simle{%  ``less than about'' symbol
    \mathrel{\rlap{\raise 0.511ex 
        \hbox{$<$}}{\lower 0.511ex \hbox{$\sim$}}}}

\def\slashchar#1{\setbox0=\hbox{$#1$}           % set a box for #1
   \dimen0=\wd0                                 % and get its size
   \setbox1=\hbox{/} \dimen1=\wd1               % get size of /
   \ifdim\dimen0>\dimen1                        % #1 is bigger
      \rlap{\hbox to \dimen0{\hfil/\hfil}}      % so center / in box
      #1                                        % and print #1
   \else                                        % / is bigger
      \rlap{\hbox to \dimen1{\hfil$#1$\hfil}}   % so center #1
      /                                         % and print /
   \fi}

\def\sigsi{\sigma_{SI}}
\def\sigsd{\sigma_{SD}}
\def\nmhd{NMHDECAY}
\def\cogent{CoGeNT}
\def\asoft{A_{soft}}

\begin{document}

%\title{Light WIMP Dark Matter in Light of CoGeNT \\or\\ What If CoGeNT, DAMA and CDMS Are Seeing Dark Matter?}

\title{CoGeNT, DAMA, and Neutralino Dark Matter in the Next-To-Minimal Supersymmetric Standard Model}
\author{John F.~Gunion$^{1}$, Alexander V.~Belikov$^2$, and Dan Hooper$^{3,4}$}

\affiliation{$^1$Department of Physics, University of California,
  Davis, CA 95616\\${^2}$ Department of Physics, The University of
  Chicago, Chicago, IL 60637 \\ $^3$ Center For Particle Astrophysics,
  Fermi National Accelerator Laboratory, Batavia, IL 60510\\ $^4$
  Department of Astronomy and Astrophysics, University of Chicago,
  Chicago, IL 60637}

\begin{abstract}

  We assess the extent to which the NMSSM can allow for light dark
  matter in the $2\gev\lsim \mcnone\lsim 12\gev$ mass range with
  correct relic density and large spin-independent direct-detection
  cross section, $\sigsi$, in the range suggested by \cogent\ and DAMA.  For
  standard assumptions regarding nucleon $s$-quark content and
  cosmological relic density, $\rho$, we find that the NMSSM falls short by a
  factor of about 10 to 15 (3 to 5) without (with) significant violation of the
  current $(g-2)_\mu$ constraints.

\end{abstract}

\pacs{95.35.+d, 12.60.Jv, 14.80.Da; UCD-HEP-TH-2010-13; FERMILAB-PUB-10-362-A }

\maketitle

%%%%%%%%%%%%%%%%%%%%%%%%%%%%%%%%%%%%%%%%%%%%%%%%%%%%%%%%%%%%%%%%%%%%%%

\section{Introduction}

The CoGeNT collaboration has announced detection of very low energy
events which are not consistent with any known
backgrounds~\cite{cogentnew}. One possible interpretation of these
events is elastic scattering of a light dark matter particle ($m\sim
5-10$ GeV) with a spin-independent cross section, $\sigsi$, on the
order of $2\times 10^{-40}$ cm$^2$ (\ie\ $2\times
10^{-4}\pb$)~\cite{cogentnew,liam,Kopp:2009qt,Chang:2010yk}. This is
not very far from the region required to explain the annual modulation
observed by the DAMA/LIBRA collaboration~\cite{DAMAnew}. A consistent
interpretation of both the DAMA and \cogent\
observations~\cite{Hooper:2010uy} is for dark matter to have mass and
cross section in a $2\sigma$ ellipse ranging from $\sigsi\sim 3\times
10^{-4}\pb$ at $m\sim 6\gev$ down to $\sigsi \sim 1.4\times
10^{-4}\pb$ at $m\sim 9\gev$. Clearly, it is of great interest to
explore different kinds of dark matter models with regard to their
ability to yield large $\sigsi$ for $m\sim 6-9\gev$.

A number of groups have addressed this issue within the context of the
minimal supersymmetric standard model
(MSSM)~\cite{Feldman:2010ke,Kuflik:2010ah}. However, given the
structure of the MSSM Higgs sector and constraints thereon from LEP
and elsewhere, achieving the above cross section at low LSP mass is
not possible~\cite{lightLSP,Feldman:2010ke}.  A much higher local
density of dark matter than the measured cosmological dark matter
density, $\rho=0.3$~GeV/cm$^3$, would be needed to bring the $\sigsi$
required to describe the \cogent/DAMA events down to the level possible
within the MSSM.  Basically, the problem is that the Higgs with
Enhanced coupling to down quarks, whose exchange is primarily
responsible for the elastic scattering of the LSP (the lightest
neutralino) on a nucleon, must be rather heavy in the MSSM context
after imposing LEP constraints. Of course, a local density much larger
than the cosmological average could be assumed so as to get the needed
$\sigsi$ at low $m$. However, there is a second problem.  For low LSP
mass, the MSSM simply does not allow sufficient early universe
annihilation to yield the observed cosmological average relic density
once Tevatron limits on $\br(B_s\to \mupmum)$ are
imposed~\cite{Feldman:2010ke}.

Thus, it is interesting to see if an extension of the MSSM could allow
the relevant Higgs boson to have lower mass than allowed in the MSSM,
thereby achieving $\sigsi= (1.4-3.5)\times 10^{-4}\pb$, while maintaining
consistency with all constraints.  In a previous
paper~\cite{Belikov:2010yi}, we explored this question within the
context of supersymmetric models with an additional generic chiral
singlet superfield and found that this was indeed possible, the
successful scenarios being ones in which both the LSP and exchanged
Higgs are substantially singlet in nature.  In this paper, we focus on
the concrete (and more restrictive) case of the next-to-minimal
supersymmetric standard model (NMSSM). Our conclusion will be that the
observed cosmological relic density can be achieved while maintaining
consistency with limits on $\br(B_s\to \mupmum)$ but that the largest
$\sigsi$ values that can be achieved for standard inputs regarding the
$s$-quark content of the nucleon fall short of the preferred $\sigsi$
region of \cite{Hooper:2010uy} by a significant factor.
In particular, in the strict
NMSSM, scenarios with a light singlet $\cnone$ and largely singlet
light Higgs cannot be realized at high $\tanb$ while satisfying all
other constraints. We also briefly discuss possibilities for enhancing
the NMSSM cross sections by enhancing the $s$-quark nucleon content or
reducing the required $\sigsi$ using the recently proposed larger
local density $\rho\sim [0.4-0.485]$~GeV/cm$^3$ (see
\cite{Pato:2010yq} for a summary).

The remainder of this paper is structured as follows. In
Sec.~\ref{mssm}, we outline the problems faced in the MSSM. In
Sec.~\ref{extended}, we discuss how the NMSSM can potentially avoid
these problems without violating the relevant collider
constraints. In Sec.~\ref{scans}, we turn to a detailed discussion of
the NMSSM, including the point searching procedures we will employ and
the constraints that must be obeyed. In Sec.~\ref{nmssm}, we present
the NMSSM benchmark points we have found with large $\sigsi$ that
satisfy all LEP and BaBar limits.  We then examine implications of
various additional constraints from the Tevatron, $B$ physics and
$(g-2)_\mu$ for such points. We discuss some phenomenological issues
for those points that survive all constraints.  In
Sec.~\ref{conclusions}, we summarize our results and draw conclusions.

\section{Light Neutralinos In The MSSM}
\label{mssm}

In the MSSM, there are two CP-even Higgs bosons, the $\hl$ and the
$\hh$ with $\mhl<\mhh$.  In the usual convention, one writes
$\hh=\cos\alpha H_d+\sin\alpha H_u$, $\hl=-\sin\alpha H_d +\cos\alpha
H_u$, where $H_{d,u}$ are the neutral Higgs fields that couple to down
and up type quarks respectively. An especially crucial parameter of
the model is $\tanb\equiv \vev{H_u}/\vev{H_d}$. Relative to the SM
Higgs, $g_{\hl VV}=\sin(\beta-\alpha)$ and $g_{\hh
  VV}=\cos(\beta-\alpha)$, where $VV=W^+W^-$ or $ZZ$. The structure of
the model combined with LEP constraints require that
$\mhl,\mhh>90-100\gev$.  In this case, $\cos(\beta-\alpha)$ must be
fairly small, especially at large $\tanb$. The combination of large
$\tanb$ and small $\cos(\beta-\alpha)$ implies $\alpha\sim 0$ and
$\cos\alpha\sim 1$. In this situation, the only way to get a large
spin-independent cross section for lightest neutralino, $\cnone$,
scattering on the nucleon is via exchange of the $\hh$ between the
$\cnone$ ($g_{\hh \cnone\cnone}\propto \cos\alpha$) and the down type
quarks contained in the nucleon ($g_{\hh
  dd,ss,bb}\propto\tanb\cos\alpha$).  A rough formula for the
spin-independent cross section was obtained in
\cite{Belikov:2010yi}:
\begin{eqnarray}
\sigsi
&\approx& 1.7 \times 10^{-5}\pb \, \bigg(\frac{N^2_{13}}{0.1}\bigg) \bigg(\frac{\tan \beta}{50}\bigg)^2 \bigg(\frac{100 {\rm GeV}}{\mhh}\bigg)^4\cos^4\alpha\,,
\end{eqnarray}
where we have written $\cnone=N_{11}\wtil B+ N_{12} \wtil
W^3+N_{13}\wtil H_d+N_{14}\wtil H_u$.  In the above, $N_{13}^2$ cannot
be much larger than $0.1$ because of limits on the $Z$ invisible
width.  Given that LEP constraints basically force $\mhh\gsim 100\gev$
and that other constraints (including $b$-quark Yukawa perturbativity)
are very difficult to satisfy for $\tanb\geq 50$, 
we see that the MSSM is unable to obey all
constraints and yield $\sigsi$ larger than a fraction of $10^{-4}
\pb$. 

In addition, one must consider whether the MSSM allows for
sufficient early-universe annihilation to achieve 
$\Omega_{\cnone}h^2 < 0.1$. To  briefly review, 
the density of neutralino dark matter in the universe today can be
determined by the particle's annihilation cross section and mass. In
the mass range we are considering here, the dominant annihilation
channel is to $b\bar{b}$ (or to a lesser extent to $\tau^+ \tau^-$)
through the $s$-channel exchange of the pseudoscalar Higgs boson,
$A$. The thermally averaged cross sections for these processes are
given by
\begin{eqnarray}
\vev{\sigma_{\cnone\cnone \rightarrow A \rightarrow b\bar{b},\tauptaum }\, v} &=& \frac{(3,1) g^2_2 m^2_{b,\tau} \tan^2\beta}{8 \pi m^2_W}\frac{m^2_{\cnone} \sqrt{1-m^2_{b,\tau}/m^2_{\cnone}}}{(4 m^2_{\cnone}-\mha^2)^2 + m^2_A \Gamma^2_{A^0}} \nonumber \\
&\times& [(N_{13} \sin \beta - N_{14} \cos \beta) (g_2 N_{12}- g_1 N_{11})]^2,
\end{eqnarray}
where $\Gamma_{A^0}$ is the width of the pseudoscalar MSSM Higgs. And although
there are additional contributions from scalar Higgs exchange, these
are suppressed by the square of the relative velocity of the
neutralinos, and thus are substantially suppressed in the process of
thermal freeze-out.

The thermal relic abundance of neutralinos is given by
\begin{equation}
\Omega_{\cnone} h^2 \approx \frac{10^9}{M_{\rm Pl}}\frac{m_{\cnone}}{T_{\rm FO} \sqrt{g_{\star}}}\frac{1}{\vev{\sigma_{\cnone \cnone} v}}
\end{equation}
where $g_{\star}$ is the number of relativistic degrees of freedom available at freeze-out and $T_{\rm FO}$ is the temperature at which freeze-out occurs:
\begin{equation}
\frac{m_{\cnone}}{T_{\rm FO}} \approx \ln\bigg(\sqrt{\frac{45}{8}}\frac{m_{\cnone} M_{\rm Pl} \, \vev{\sigma_{\cnone \cnone} v}}{\pi^3 \sqrt{g_{\star} m_{\cnone}/T_{\rm FO}}} \bigg). 
\end{equation}
For the range of masses considered here, and for cross sections which will yield approximately the measured dark matter abundance, we find $m_{\cnone}/T_{\rm FO}\approx 20$.

For $m_{\cnone}\sim 5-15$ GeV, the relic abundance of MSSM neutralinos
is then approximately given by
\begin{eqnarray}
\Omega_{\cnone} h^2 \approx 0.1 \, \bigg(\frac{0.1}{N^2_{13}}\bigg) \bigg(\frac{50}{\tan \beta}\bigg)^2 \bigg(\frac{\mha}{100 \, {\rm GeV}}\bigg)^4 \bigg(\frac{9 \, {\rm GeV}}{m_{\cnone}}\bigg)^2.
\end{eqnarray}
Given that LEP limits require $\mha\gsim 90-100\gev$ and that $\tanb$
as large as $50$ is already in the non-perturbative domain for the
$b$-quark coupling, it requires a very extreme choice of parameters to
get the measured dark matter density of our universe to be as small as
that measured, $\Omega_{\rm CDM} h^2 = 0.1131 \pm
0.0042$~\cite{wmap}. And, even with such extreme parameter choices,
$\sigsi$ can be no larger than $\sim 1.7\times 10^{-5}\pb$. Of course,
it is true that the same extreme choice of parameters that minimizes
$\Omega_{\cnone}h^2$, bringing it close to the observed value, at the
same time maximizes $\sigsi$. However, there is a further barrier to
achieving the minimal $\Omega_{\cnone}h^2$, maximal $\sigsi$ scenario.
In particular, the above discussion does not yet include consideration
of the Tevatron limits on $\br(B_s\to \mupmum)$.  In
\cite{Feldman:2010ke} (see their Fig.~3b), it was found that the MSSM
simply cannot give the correct relic density for $\mcnone$ in the
\cogent/DAMA region once the $\br(B_s\to\mupmum)$ limit is imposed in
addition to the LEP limits.  This situation motivates us to consider
supersymmetric scenarios beyond the MSSM. In the next section, we will
demonstrate that in the NMSSM it is possible to alleviate both the
elastic scattering cross section and relic abundance problems found in
the MSSM.

\section{The NMSSM}
\label{extended}

In the NMSSM, one adds exactly one singlet chiral superfield to the
MSSM. As is well known, this allows a completely natural
explanation for the size of the $\mu$ term
\cite{nmssm} and can reduce electroweak
fine-tuning \cite{Dermisek:2005ar}, and potentially catalyze
electroweak baryogenesis \cite{Funakubo:2002yb}.  The NMSSM
superpotential is given by
\begin{eqnarray}
 \lambda \hat{S} \hat{H}_u \hat{H}_d
+ \third \kappa \hat{S}^3 ~,
\label{eq:W}
\end{eqnarray}
and the associated part of the soft Lagrangian is given by 
\begin{eqnarray}
\lambda A_\lambda S H_u H_d +  \third\kappa A_\kappa S^3 + H.c.
\label{eq:Lsoft}
\end{eqnarray}
The restriction to the forms given above is implemented by invoking a
$Z_3$ symmetry to remove all other possible terms. In particular, only
the dimensionless $\lam$ and $\kap$ superpotential terms are allowed.
All dimensionful parameters are generated by soft-SUSY-breaking.  An
effective $\mu$ value is automatically obtained as $\mueff=\lam
\vev{S}$.  This very attractive extension of the MSSM allows for a
considerable expansion of the phenomenological possibilities. In
particular, the singlet superfield leads to five neutralinos, three
CP-even Higgs bosons ($h_{1,2,3}$) and two CP-odd Higgs bosons
($a_{1,2}$). In general, the neutralino mass eigenstates are mixtures
of the MSSM neutralino fields and the singlino field that is
part of the singlet superfield; the CP-even (odd) Higgs mass
eigenstates are similarly mixtures of the CP-even (odd) MSSM fields
and the CP-even (odd) components of the complex singlet scalar
component of the singlet superfield.

Within the NMSSM, it is very natural for the lightest pseudoscalar
Higgs, $\ai$, to have low mass (see~\cite{Dermisek:2006wr}). In
particular, $U(1)_R$ or $U(1)_{\rm PQ}$ symmetries can appear which
lead to values of $\mai$ well below the electroweak scale. If one is
close to either symmetry limit, the $\ai$ will be at least moderately
singlet-like (as opposed to being more purely MSSM-Higgs-like) and
will likely be beyond the reach of current collider constraints.

That a light $\ai$ in the NMSSM can allow a very light dark matter
particle in the \cogent\ mass region with correct relic density was
established in~\cite{gunion}. This is because the light $\ai$
$s$-channel annihilation process is typically fairly close to being
'on-pole', $2\mcnone\sim \mai$, as opposed to $2\mcnone\ll \mha$ for
the rather heavy $\ha$ of the MSSM.  However, in the scans performed
in \cite{gunion} we did not encounter points with cross sections as
large as those needed to describe the tentative \cogent/DAMA signal.  We
now describe a strategy for getting the largest possible cross
section.

To enhance the neutralino's elastic scattering cross section, we need
a Higgs mass eigenstate that is primarily $H_d$ (so that it will have
enhanced couplings to down-type quarks at large $\tanb$) with mass
lower than possible for the $\hh$ of the MSSM. While this is not as easy
to arrange in the NMSSM as are low values of the lightest CP-odd Higgs
mass, it is still possible. The value of the down-type diagonal term
of the NMSSM scalar Higgs (squared) mass matrix at tree-level is given
by
\begin{equation}
m^2_{H,22} = \frac{g^2 v^2}{1+\tan^2\beta} + \mu \tan \beta (A_{\lambda}+\kappa \mu /\lambda),
\end{equation}
where $v$ is the Standard Model Higgs vacuum expectation value. At
large $\tanb$, in
order for this to fall significantly below the value of the up-type
Higgs entry (which is generally $m^2_{H,11} \approx (85 {\rm
  GeV})^2)$, there must be some cancellation between the 
$A_{\lambda}$ and  $\kappa \mu/\lambda$ terms. This cancellation also
suppresses the mixing term between up-type and down-type scalar Higgs
bosons. The down-type mass can further be protected from large radiative
corrections if the two stop masses are similar. Together, these
features can potentially lead to a down-type scalar Higgs boson with a
mass significantly below $100\gev$.  

The scenarios that can potentially lead to large $\sigsi$ are then
ones in which the lightest of the NMSSM Higgs bosons, the $\hi$, is
not SM-like, has enhanced coupling to down-type quarks and has mass
below $\sim 100\gev$. The $\hii$ will typically be SM-like and for 
$\mhi$ below $100\gev$ is typically not very heavy -- $\mhii\gsim
110\gev$ for $\msusy=500\gev$ and $\mhii\gsim 115\gev$ for
$\msusy=1\tev$. LEP limits will be very
constraining in this situation. In addition, many $B$-physics
constraints will enter as will constraints from $(g-2)_\mu$.  Also
important will be limits on $b\anti b +Higgs$ production with
$Higgs\to\tauptaum$ and $t\to \hp b$ decays with $\hp\to
\tau^+\nu_\tau$. We will employ augmented versions of
NMHDECAY~\cite{Ellwanger:2004xm,Ellwanger:2005dv} supplemented by
micrOMEGAs~\cite{Belanger:2006is} (the latter will be implemented as in
NMSSMTools~\cite{nmssmtools}) for our exploration of the NMSSM
parameter space.

\section{Constraints and Scanning in the NMSSM}
\label{scans}

As noted, we have performed our scanning using an augmented version of
NMHDECAY linked to micrOMEGAs as in NMSSMTools. NMHDECAY currently
incorporates all LEP limits on Higgs bosons as well as LEP limits on
neutralinos and charginos.\footnote{We have retained the stronger
  cross section constraints of the original NMHDECAY program rather
  than weakening them in the manner suggested
  in~\cite{Das:2010ww}. However, we have updated the limit on
  $\Gamma_{Z\to \cnone\cnone}$ to $1.9\mev$ as in \cite{Das:2010ww}.}
We have augmented \nmhd\ to include the recent ALEPH
constraints~\cite{Schael:2010aw} on $\epem\to Z +Higgs$ with $Higgs\to
aa$ (in our case $a=\ai$, the lightest CP-odd Higgs boson of the
NMSSM) with $a\to \tauptaum$. Further, we have augmented NMHDECAY to
include the combined CDF+D0 Tevatron
constraints~\cite{Benjamin:2010xb} on $b\anti b + Higgs$ production
with $Higgs\to \tauptaum$ (for the scans performed in this paper, it
is constraints in the case of $Higgs=\hi$ or $\aii$ that are typically
relevant).~\footnote{Experimental plots assume the MSSM for which the
  $H$ and $A$ are nearly degenerate whereas in most NMSSM cases $\hii$
  and $\aii$ are not degenerate, implying a somewhat weaker constraint
  on the separate $ b\anti b \hii$ and $b\anti b \aii$ couplings.}
Finally, in the scenarios with large $\sigsi$ the $\hp$ is inevitably
light enough that $t\to \hp b$ decays will be present and, since
$\tanb$ is large, $\hp \to \tau^+\nu_\tau$ will be completely
dominant.  We have thus augmented NMHDECAY to include the current D0
limits~\cite{:2009zh} on $\br(t\to \hp
b)\times\br(\hp\to\tau^+\nu_\tau)$.\footnote{Limits in this channel
  from CDF are not currently available.}  \nmhd\ also includes
analysis of a large selection of $B$ physics constraints. For our
purposes, the most important ones turn out to be $B_s\to \mupmum$,
$B^+\to \tau^+\nu_\tau$, and $b\to s\gam$.  We have also augmented
\nmhd\ to incorporate full BaBar constraints on $\Upsilon_{nS}\to \gam
a$ with $a\to \mupmum$ or $a\to \tauptaum$ as implemented
in~\cite{Dermisek:2010mg}. Finally, we have examined the \nmhd\
predictions for $(g-2)_\mu$ for high-$\sigsi$ cases.
In our search for desirable points, we have demanded that all the LEP
limits, including the ALEPH limits, are strictly obeyed. We have also
demanded that the BaBar limits be strictly satisfied.

The $b\anti b+Higgs(\to \tauptaum)$
and $t\to \hp(\to \tau^+\nu) b$ limits are treated somewhat
differently. In the experimental papers, the observed limits are
plotted as a function of the relevant Higgs mass in comparison to the
expected limits.  The expected limits have error bars that are partly
statistical and partly systematic (including theory systematics) that
have been combined in quadrature, \ie\ assuming a Gaussian
distribution in particular for theoretical systematics. We believe
that treating the observed limits in these cases as true limits is
somewhat dubious.  In our opinion, it would be much better to have separated the
statistical errors from the systematic errors and ask what band about
the observed limits would result from pushing all systematics in the
least or most favorable direction. In the absence of sufficient
information to carry out this task, we will simply assess the impact
of relaxing the observed limits in the above channels by an amount
equivalent to the $1\sig$ or $2\sig$ error bands (as plotted relative
to the expected limits) relative to the
observed limits.\footnote{In the $t\to \hp b$ case, plots only show
a  $1\sig$ error band.  We have simply doubled this for an
approximation to the $2\sig$ error band.}

In assessing the $B^+\to \tau^+\nu_\tau$, $b\to s\gam$ and $(g-2)_\mu$
constraints contained in the basic \nmhd\ program ($B_s\to\mupmum$ is
handled differently as described later) we have adopted the following
procedure. The \nmhd\ output gives the model point prediction as well
as the maximum and minimum values after adding and subtracting the
theoretical error. Let us call these $P_0$, $P_+$ and $P_-$,
respectively. Also contained in the output is the $\pm 2\sigma$
interval for the experimentally observed value or limit, which we
label as $O_{+2\sig}$ and $O_{-2\sig}$, respectively. Any point for
which $P_+$ or $P_-$ falls within the interval
$I=[O_{-2\sig},O_{+2\sig}]$ is deemed acceptable. If this is not the
case we assess the extent of the violation of the constraint as
follows.  Let us say $P_->O_{+2\sig}$. Define
$\Delta=|P_--O_{+2\sig}|$. We then compute $R_\sig=\Delta/E$, where
$E$ is a combined error associated with the experimental and
theoretical errors: $E\equiv
[(|O_{+2\sig}-O_{-2\sig}|/4)^2+(|P_+-P_-|/2)^2]^{1/2}$. If $P_+$ or
$P_-$ falls within the interval $I=[O_{-2\sig},O_{+2\sig}]$ we set
$R_\sig=0$. We will summarize the values found for $R_\sig$ for
high-$\sigsi$ points for each of the above three constraints.

In our scans, we have held fixed the soft scales $M_2=200\gev$ and
$M_3=300\gev$, allowing for varying values of $M_1$ (which essentially
fixes the mass of the bino-like neutralino).  Our scans have been
performed for fixed values of $\mueff=+200\gev$ and $-200\gev$. (It
seems that smaller $|\mueff|$ values do not allow large $\sigsi$ to be
consistent with all other constraints. Conversely, larger $|\mueff|$ tends to
lower the achievable $\sigsi$.) We have considered three values of
$\tanb$, $\tanb=40$, $\tanb=45$ (only for $\mueff<0$) and
$\tanb=50$. We have adopted a universal value of $\msusy$ for all the
soft SUSY-scale slepton and squark SUSY-breaking masses.  We consider
$\msusy=500\gev$ and $1\tev$. We have adopted a universal value for
all the soft $A$ parameters, \ie\ $\asoft\equiv A_t=A_b=A_\tau,\ldots$.  It
turns out that essentially the only way to obtain a value for
$\br(B_s\to \mupmum)$ below the current experimental limit when
$\tanb$ is large  is to
choose $\asoft$ rather precisely (typically to within 1\%). At high
$\tanb$, it turns out that the appropriate choice for $\asoft$ is
essentially only a function of $\msusy$. For each choice of $\msusy$,
we have determined the appropriate $\asoft$ and have then held it fixed at
this value as we scan over other parameters and assess all the other
constraints (LEP, BaBar, Tevatron, \ldots).

In all our scans, we have consistently found that large $\sigsi$ is
only achieved if the $\cnone$ is
mostly bino, implying that $\mcnone$ is pretty much fixed to be close
to $M_1$. As a result, we have performed scans at a variety of $M_1$
values in the general \cogent\ range. For any given $M_1$ we thus end up
scanning in $\lam,\kap,\alam,\akap$, demanding, as sketched above,
complete consistency with all LEP and BaBar limits, but allowing for
some deviation from $B$-physics, Tevatron and $(g-2)_\mu$ nominal
constraints.  For a choice of $\lam,\kap,\alam,\akap$ that is allowed
by LEP and BaBar constraints (at the given $\asoft$), there is
no guarantee that $\Omega h^2\sim 0.11$ will be obtained.
Fortunately, it is often the case that one can adjust $\mai$ (by
changing $\akap$ by a relatively small amount) and or $\mcnone$ (by
changing $M_1$) so that $\Omega h^2\sim 0.11$ (we accept points within
the NMSSMTools-defined window, $0.094\leq \Omega h^2\leq 0.136$) 
is achieved without destroying consistency with
LEP and BaBar limits.  The results of these scans after this
adjustment are presented in the following section.

\section{Benchmark Models In The NMSSM}
\label{nmssm}

\begin{figure}[h!]
\begin{center}
\includegraphics[width=0.65\textwidth,angle=90]{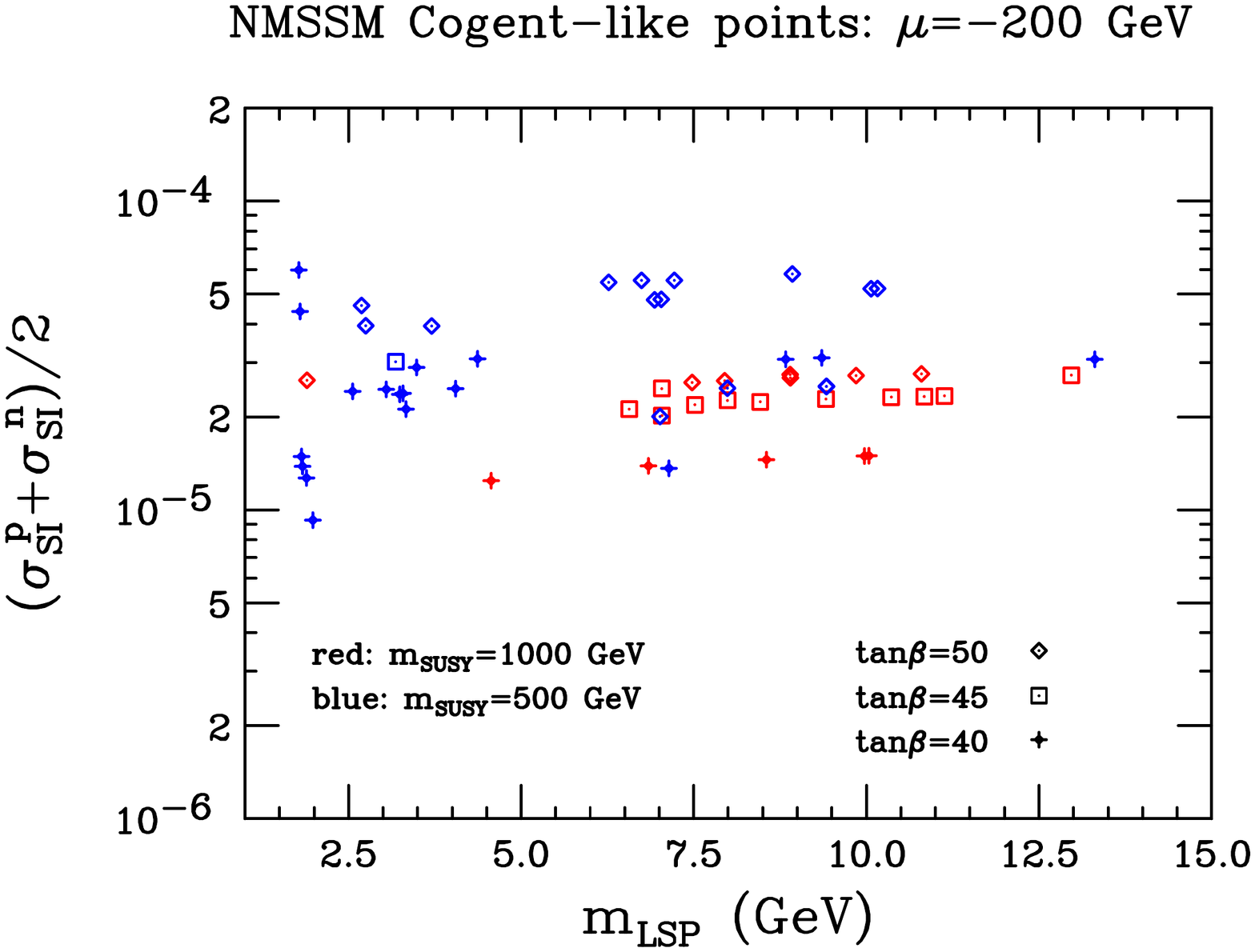}
\end{center}
\caption{$\sigsi$ vs. $\mcnone$ for $\mueff=-200\gev$. Parameters not shown are fixed as stated in the text. Only
level-I constraints are imposed.}
\label{mu-200allpts}
\end{figure}

\begin{figure}[h!]
\begin{center}
\includegraphics[width=0.65\textwidth,angle=90]{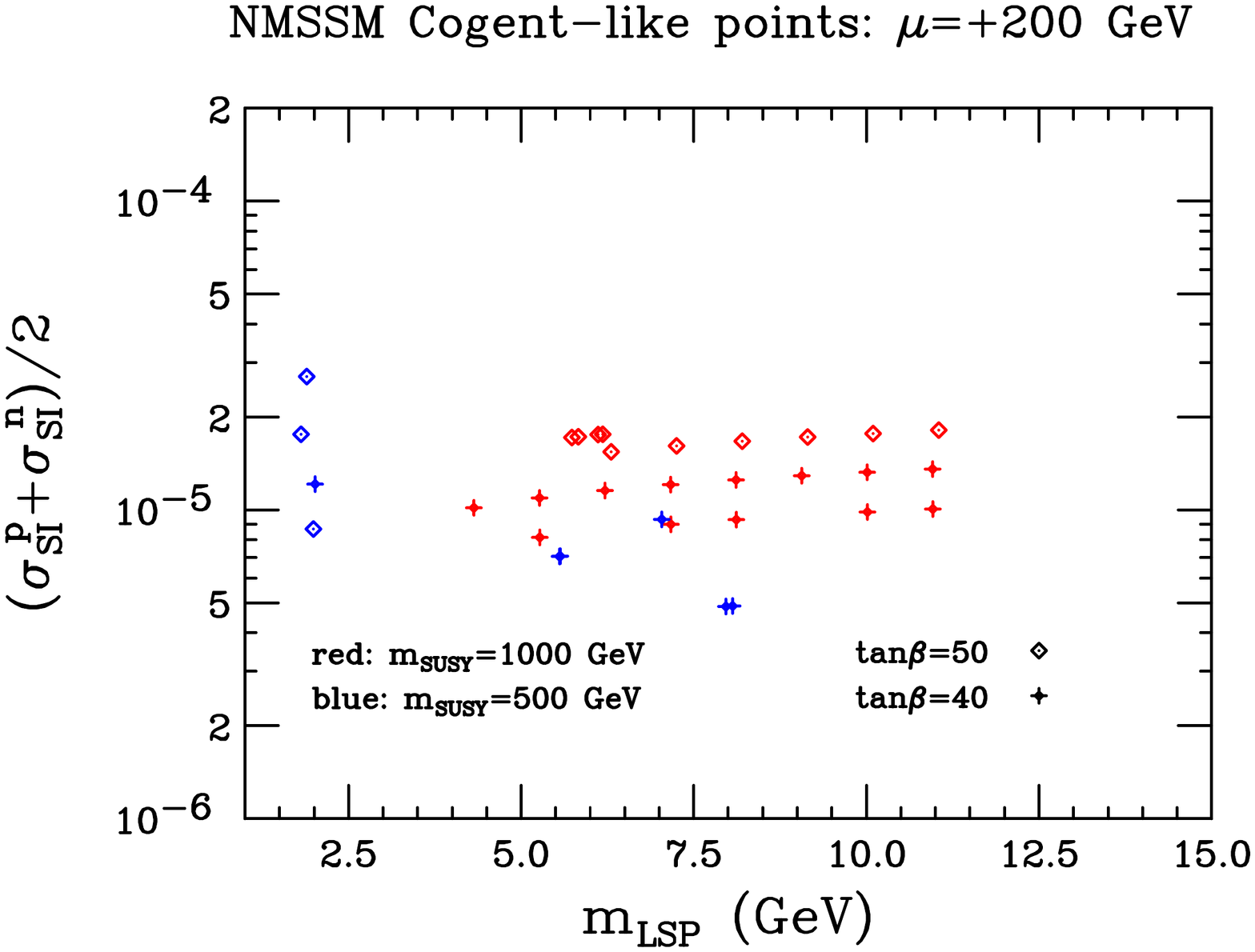}
\end{center}
\caption{$\sigsi$ vs. $\mcnone$ for $\mueff=+200\gev$. Parameters not shown are fixed as stated in the text. Only level-I constraints are imposed.}
\label{mu+200allpts}
\end{figure}

We begin with plots, Figs.~\ref{mu-200allpts} and \ref{mu+200allpts},
of $\sigsi$ vs. $\mcnone$ for $\mueff=-200\gev$ and
$\mueff=+200\gev$. We only give points found that have fairly large
$\sigsi$. For these two figures, only the LEP constraints, BaBar
constraints, $\br(B_s\to \mupmum)$ limits and $0.094\leq \Omega
h^2\leq 0.136$ are required to be satisfied. We refer to these as
level-I constraints.  Many of the plotted points with the largest $\sigsi$
values fail at some level one or more of the other constraints, as we
shall describe. 

For $\mueff=-200\gev$ we see in Fig.~\ref{mu-200allpts} that fairly
large values of $\sigsi$ (only a factor of 3 to 5 or so below the values
typical of the preferred \cogent/DAMA region) can be obtained.  Such
points typically have both large $\tanb=50$ and low $\msusy$ (so that
$\mhi$ can be relatively smaller). In contrast,
Fig.~\ref{mu+200allpts} shows that for $\mueff=+200\gev$ we never
found any points with $\tanb=50$ and $\msusy=500\gev$ that were
consistent with LEP and BaBar limits. Consistent points were found for
$\tanb=40$ and $\msusy=500\gev$ with $\sigsi\sim 0.1\times
10^{-4}\pb$. For $\msusy=1\tev$, consistent points are found for both
$\tanb=50$ and $\tanb=40$ for which the largest cross sections found
are of order $0.2\times 10^{-4}\pb$ and $0.15\times 10^{-4}\pb$,
respectively, both of which are significantly 
below the cross section needed to explain \cogent/DAMA events.

As anticipated from our earlier discussions, one finds that almost all
the high-$\sigsi$ points for either sign of $\mueff$ have $C_V(\hi)\ll
1$ (where $C_V(h)=g_{hVV}/g_{\hsm VV}$), implying that either $\hii$
or $\hiii$ is the SM-like Higgs boson. This was not imposed, but
simply came out of the scan when large $\sigsi$ was required.  This
shows that our intuition as to how to achieve large $\sigsi$ was
correct. For many cases, $\mhii<110\gev$ and $C_V(\hii)\sim 1$.
Such points escape LEP limits because $\br(\hii\to\ai\ai)$ is large
and $10\gev\lsim \mai\lsim 2m_B$, the $10\gev$ lower bound so that BaBar
constraints on $\Upsilon_{3S}\to \gam\ai$ and ALEPH constraints on
$Z\hii$ with $\hii\to\ai\ai\to 4\tau$ are obeyed and the upper bound
so that $\ai\to b\anti b$ is forbidden. 

\begin{figure}[h!]
\begin{center}
\hspace*{-.3in}\includegraphics[width=0.5\textwidth,angle=90]{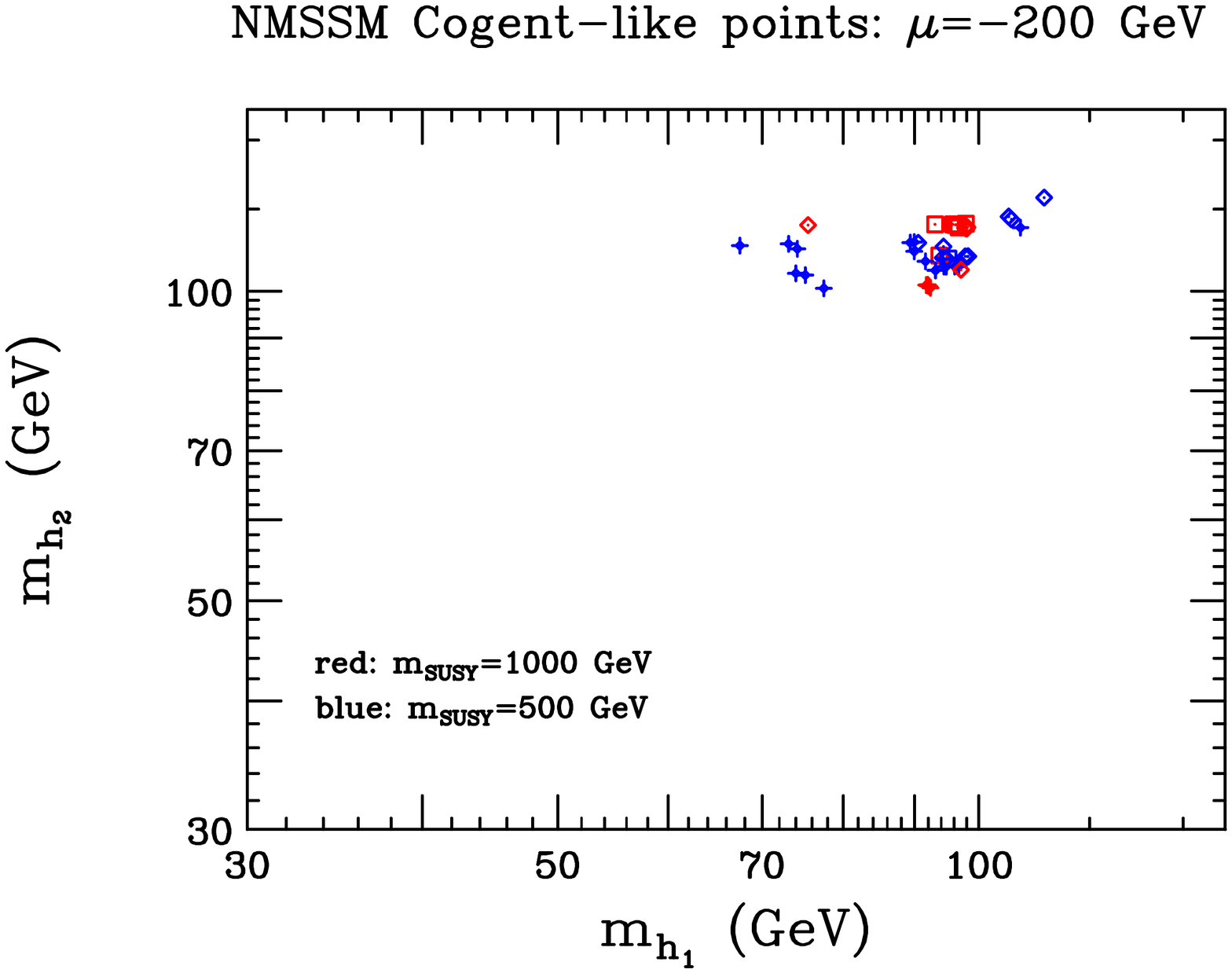}\hspace*{-1in}\includegraphics[width=0.5\textwidth,angle=90]{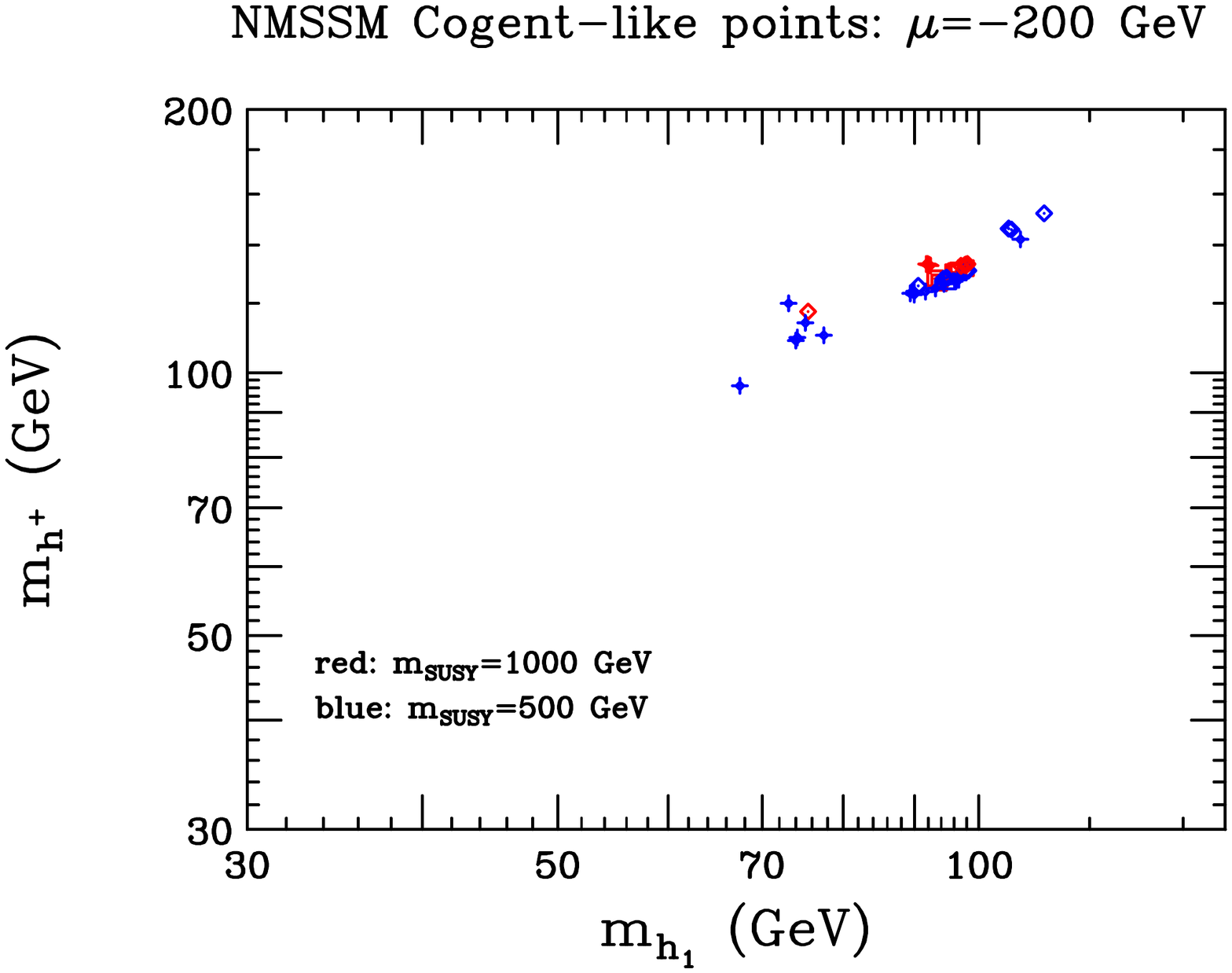}
\end{center}
\caption{$\mhii$ and $\mhp$ vs. $\mhi$ for $\mueff=-200$ points.
 Parameters not shown
  are fixed as stated in the text. Only level-I constraints are
  imposed. There is a lot of point overlap in this plot.}
\label{massesmu-200}
\end{figure}

\begin{figure}[h!]
\begin{center}
\hspace*{-.3in}\includegraphics[width=0.5\textwidth,angle=90]{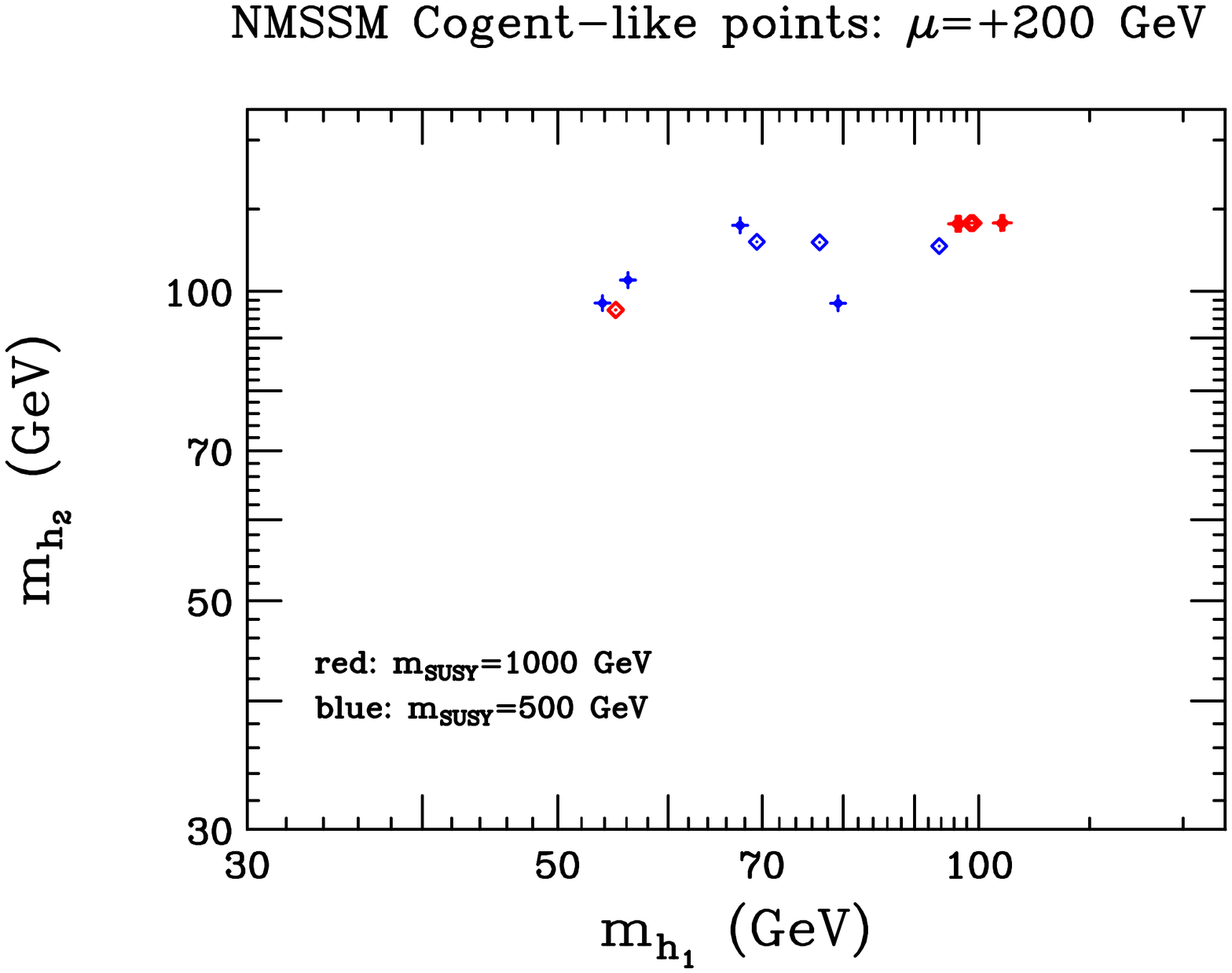}\hspace*{-1in}\includegraphics[width=0.5\textwidth,angle=90]{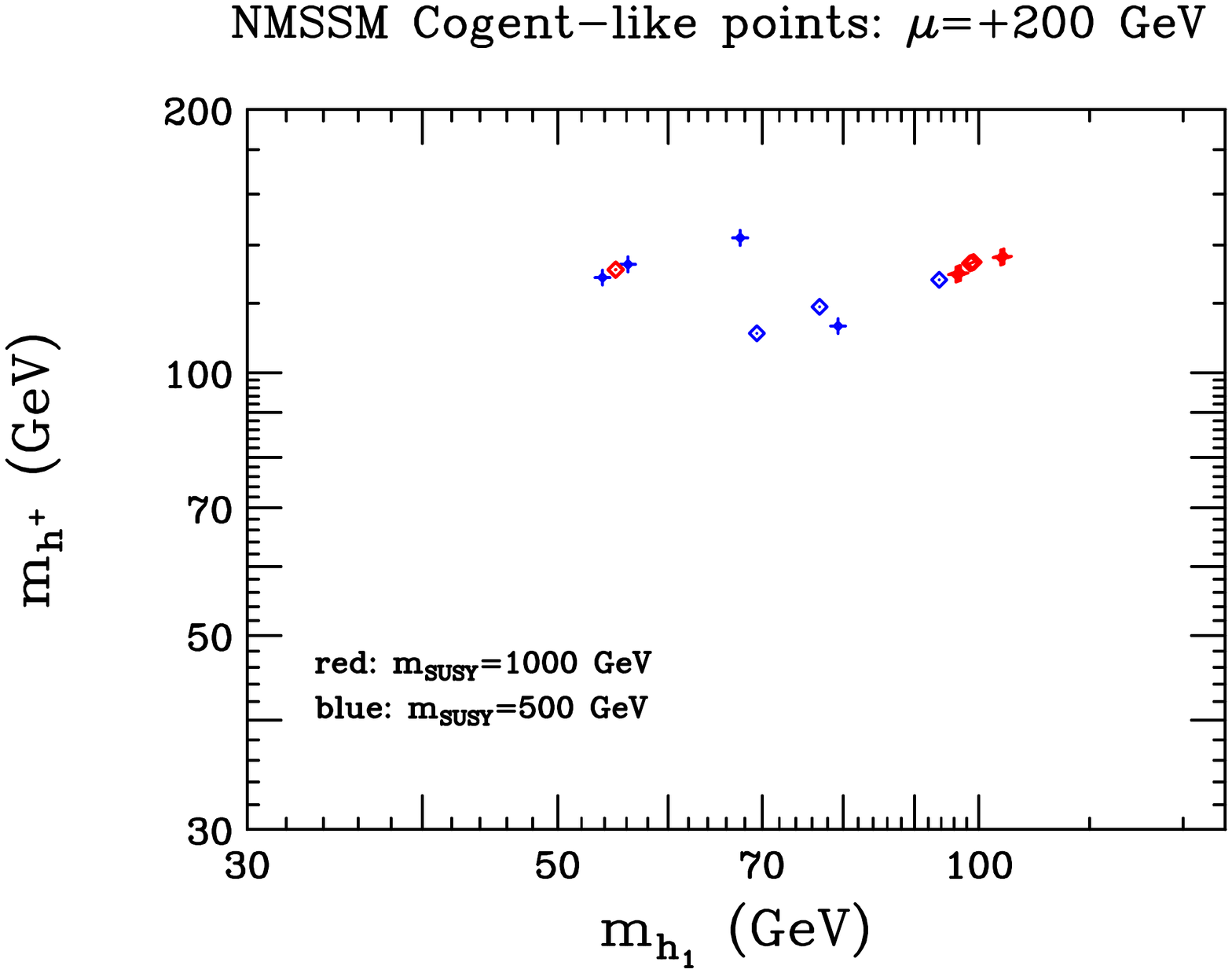}
\end{center}
\caption{$\mhii$ and $\mhp$ vs. $\mhi$ for $\mueff=+200$ points.
 Parameters not shown
  are fixed as stated in the text. Only level-I constraints are
  imposed. There is a great amount of point overlap in this plot.}
\label{massesmu+200}
\end{figure}

Of interest for the following are the masses of the $\hii$ and $\hp$
for the large $\sigsi$ points.  These are shown in
Figs.~\ref{massesmu-200} and \ref{massesmu+200}. One should take note
of the rather low values of $\mhi$, $\mhii$ and $\mhp$. (For some
points, $\mhiii$ is also quite small.)  Low $\mhp$ combined with large
$\tanb$ implies that $\br(t\to \hp b)$ will be significant and that
$\br(\hp\to \tau^+\nu_\tau)\sim 1$.  Low masses for the neutral Higgs
bosons coupled with the fact that at least several of them will have
enhanced $b\anti b $ coupling when $\tanb$ is large implies that
$b\anti b+Higgs$ with  $Higgs\to \tauptaum$ will have a high rate
at a hadron collider for several of the neutral Higgs.
Thus, Tevatron constraints will often
be of importance, and future LHC results could have a deciding impact.


\begin{figure}[h!]
\begin{center}
\includegraphics[width=0.65\textwidth,angle=90]{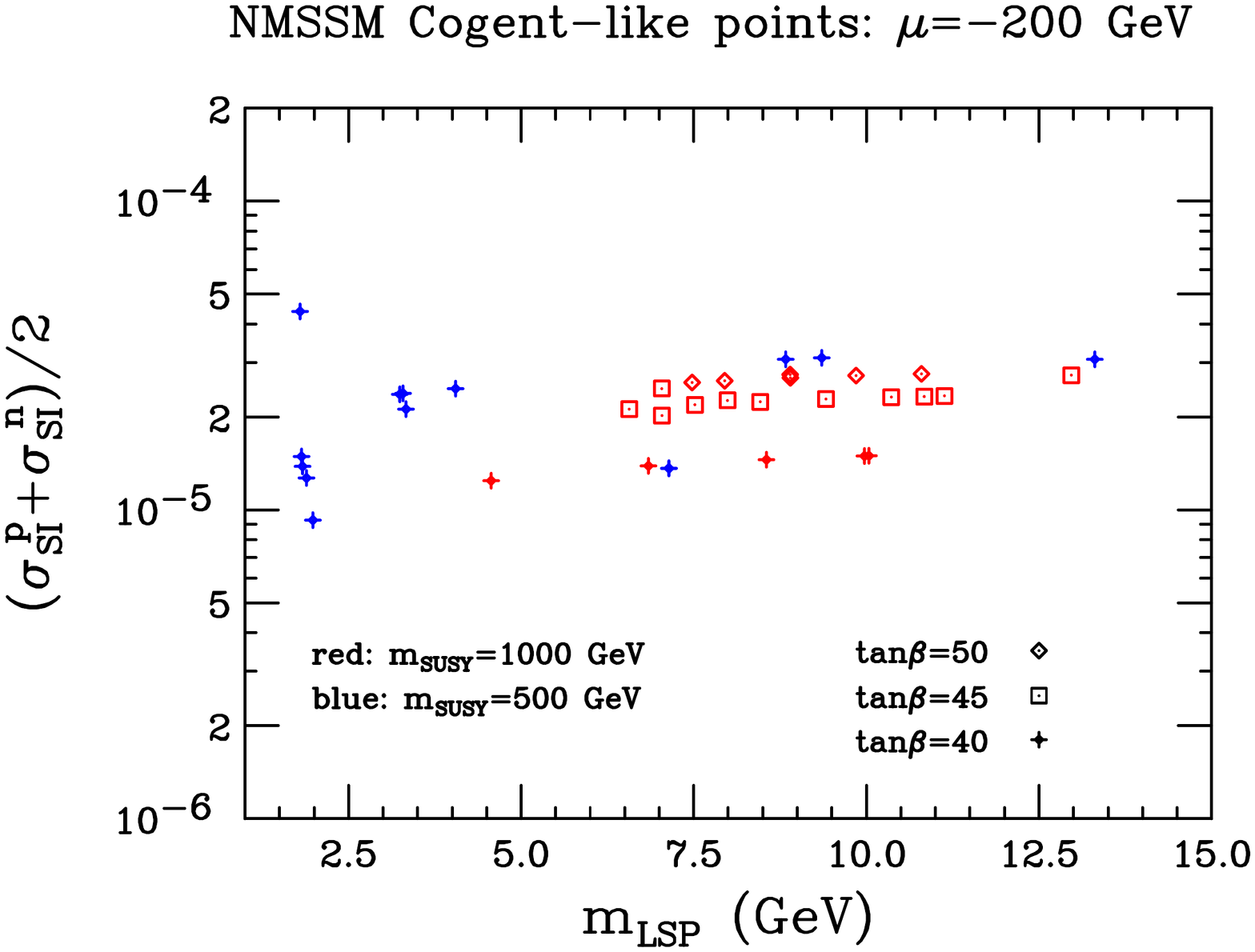}
\end{center}
\caption{$\sigsi$ vs. $\mcnone$ for $\mueff=-200\gev$ points 
  consistent within $1\sig$ (see text) with Tevatron
  limits on $b\anti b +Higgs$ and $t\to \hp b$. Parameters not shown
  are fixed as stated in the text. Level-I constraints are imposed.}
\label{mu-200_tev1sig}
\end{figure}

Indeed, let us now add to the LEP and BaBar constraints the
requirement that the Tevatron constraints on $b\anti b+Higgs$ and
$t\to \hp b$ be satisfied within $1\sig$ as defined in the previous
section.  The plot of Fig.~\ref{mu-200_tev1sig} shows that for
$\mueff=-200\gev$ the points with largest $\sigsi$ (\ie\ those with
low $\msusy$ and hence lower $\mhi$ and large $\tanb$) do not satisfy
the additional Tevatron constraints.  The maximal cross section
allowed is $\sim 0.3\times 10^{-4}\pb$, which is distinctly below the
$\sigsi=(1.4-3.5) \times 10^{-4}\pb$ of the \cogent/DAMA region.

For $\mueff=+200\gev$, the LEP and BaBar constraints had already
eliminated such points and imposing the Tevatron constraints at the
$1\sig$ level eliminates only the single point of
Fig.~\ref{mu+200allpts}  with $\mcnone\sim
2.4\gev$ and $\sigsi\sim 0.28\times 10^{-4}\pb$.

\begin{figure}[h!]
\begin{center}
\hspace*{-.3in}\includegraphics[width=0.5\textwidth,angle=90]{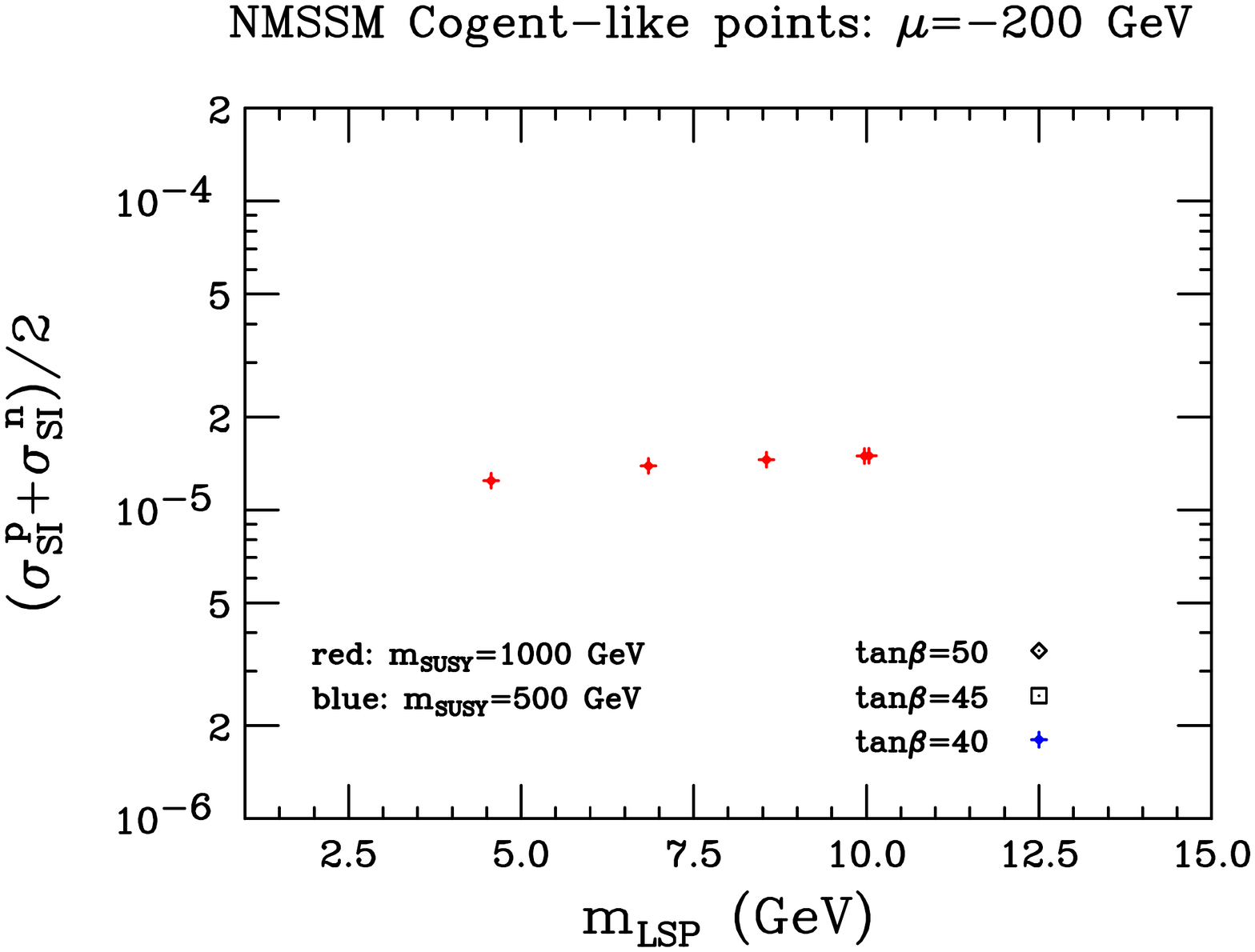}\hspace*{-1in}\includegraphics[width=0.5\textwidth,angle=90]{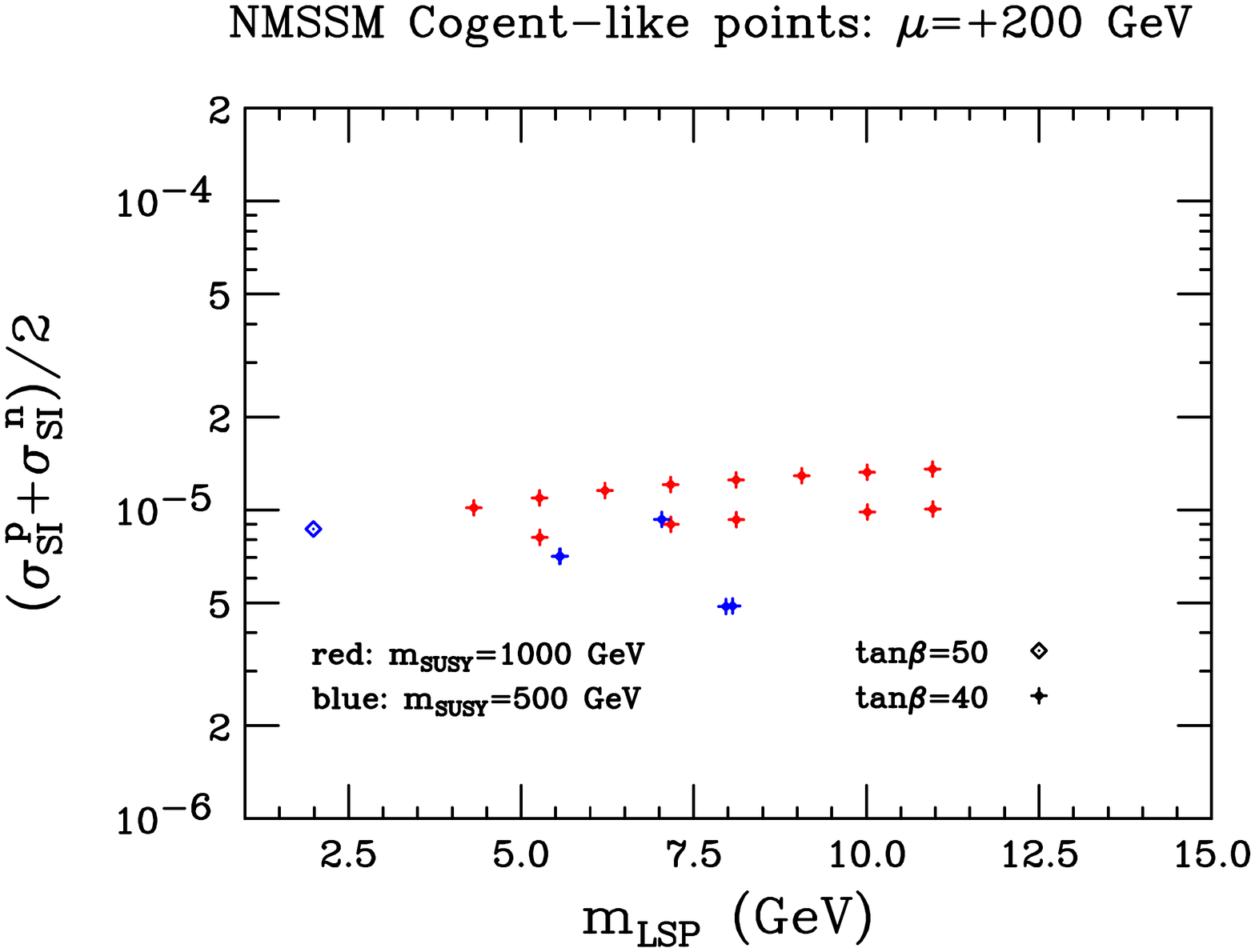}
\end{center}
\caption{$\sigsi$ vs. $\mcnone$ for points fully consistent with Tevatron
  limits on $b\anti b +Higgs$ and $t\to \hp b$. Parameters not shown
  are fixed as stated in the text. Level-I constraints are imposed.}
\label{tevok}
\end{figure}

If we require that the Tevatron observed limits apply with no
allowance for error, we obtain the plots shown in
Fig.~\ref{tevok}. The maximum $\sigsi$ for both $\mueff=-200\gev$ and
$\mueff=+200\gev$ in the \cogent\ $\mcnone$ region is of order $0.14\times
10^{-4}\pb$, a factor of $10-20$ below the $\sigsi=(1.4-3.5)\times
10^{-4}\pb$ \cogent/DAMA region.

\begin{figure}[h!]
\begin{center}
\includegraphics[width=0.5\textwidth,angle=90]{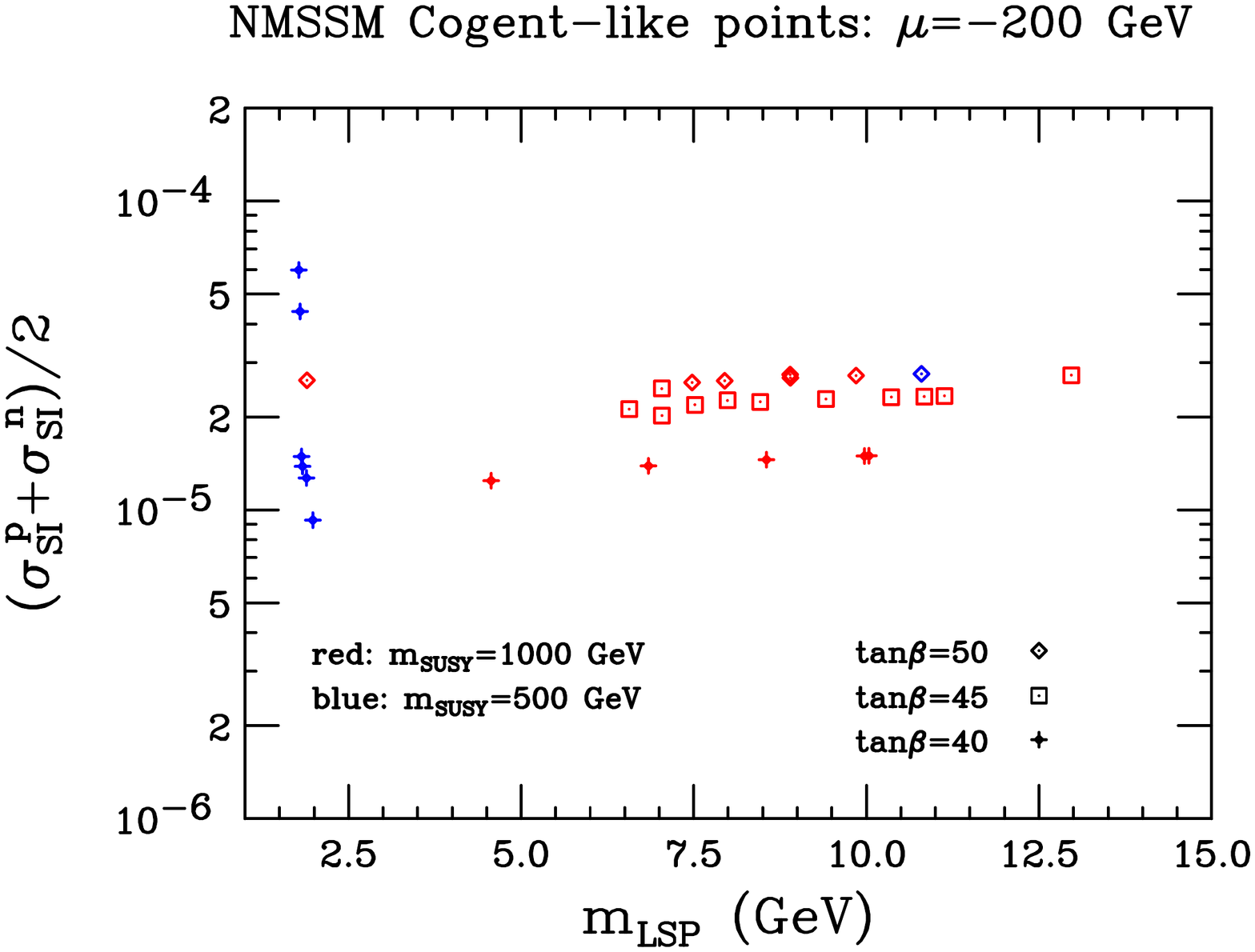}
\end{center}
\caption{$\sigsi$ vs. $\mcnone$ for $\mueff=-200\gev$ points satisfying level-I
  constraints and with $R_\sig((g-2)_\mu)<3$. Parameters not shown
  are fixed as stated in the text.}
\label{rsig3}
\end{figure}

We now turn to the impact on these results of $B^+\to \tau^+\nu_\tau$,
$b\to s\gam$ and $(g-2)_\mu$ constraints. To assess these impacts, we
employ $R_\sig$ defined earlier, where $R_\sig$ is computed for each
of the above three cases. For $\mueff=-200\gev$, non-zero values of
$R_\sig$ only arise for $B^+\to \tau^+\nu_\tau$ and $(g-2)_\mu$.
$R_\sig(B^+\to \tau^+\nu_\tau)$ for the plotted points is typically
below, often well below, $0.4$, which we do not regard as a
significant exception to the experimental limits.  On the other hand
$R_\sig((g-2)_\mu)$ is often quite large.  Indeed, if we require
$R_\sig((g-2)_\mu)<2$ then all points are eliminated except for those
with very low $\mcnone\sim 2.4\gev$.  Requiring $R_\sig((g-2)_\mu)<3$
leaves the points plotted in Fig.~\ref{rsig3}, \ie\ it is the
$\msusy=1000\gev$ points that can survive this very loose
constraint. In short, if $(g-2)_\mu$ is taken seriously, the
$\mueff=-200\gev$ points must be eliminated from consideration.  Of
course, one should never completely rule out the possibility that
significant additional new physics could contribute to $(g-2)_\mu$
without affecting the NMSSM structure of the Higgs and dark matter
sectors.

In contrast, the vast majority of the $\mueff=+200\gev$ points (and
indeed all of those near the \cogent\ mass window) are fully consistent with
both $B^+\to\tau^+\nu_\tau$ and $(g-2)_\mu$ constraints within the
\nmhd\ windows and only have small values of $R_\sig(b\to s\gam)$.
For all of the plotted points in the $\mcnone>4\gev$ region,
$R_\sig(b\to s\gam)\in [0.5,0.6]$.  Given the possibility of other new
physics that might enter into $b\to s\gam$ that might easily have no
affect on the NMSSM Higgs and dark matter issues, we regard this as
acceptable.

\begin{figure}[h!]
\begin{center}
\hspace*{-.3in}\includegraphics[width=0.5\textwidth,angle=90]{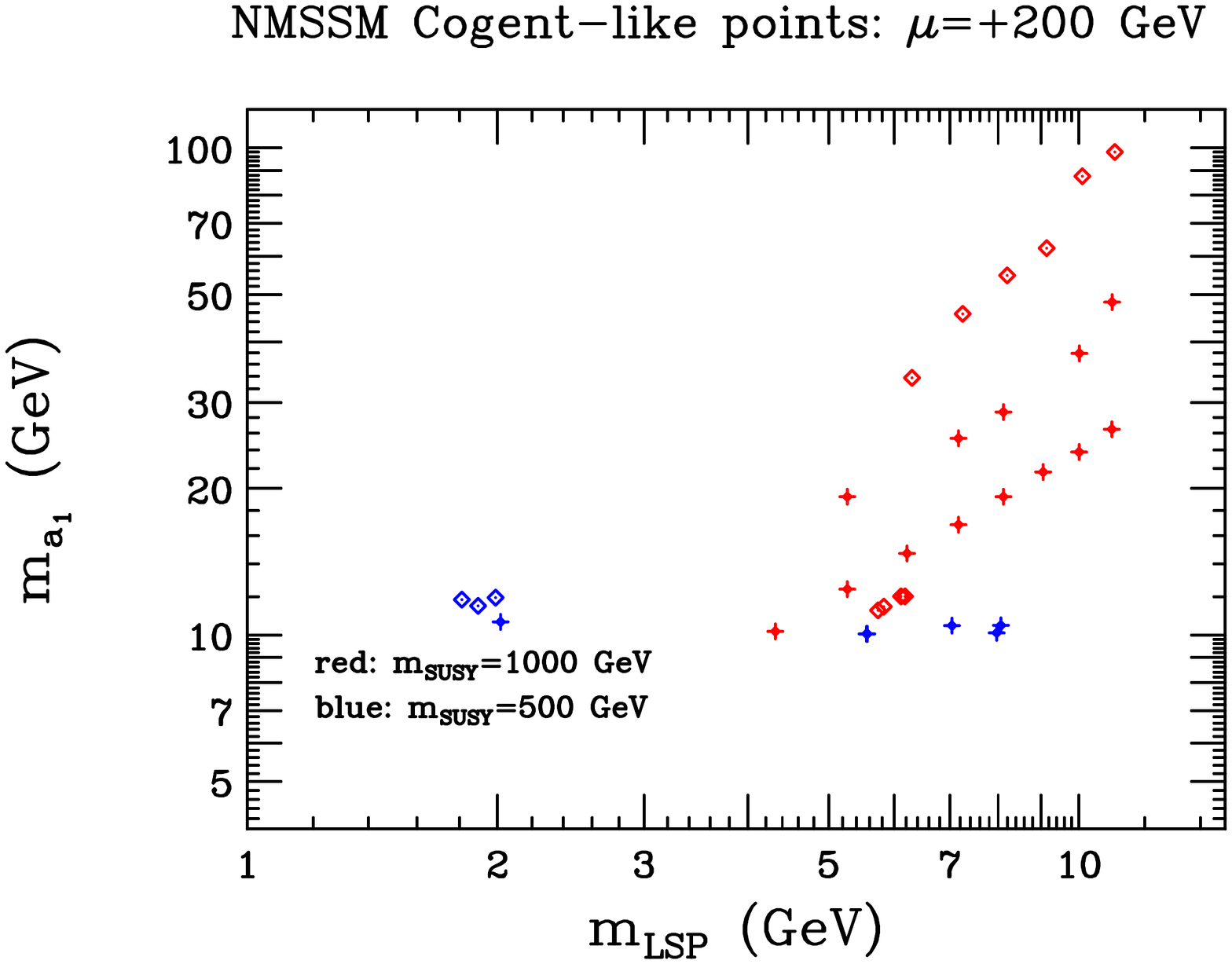}\hspace*{-1in}\includegraphics[width=0.5\textwidth,angle=90]{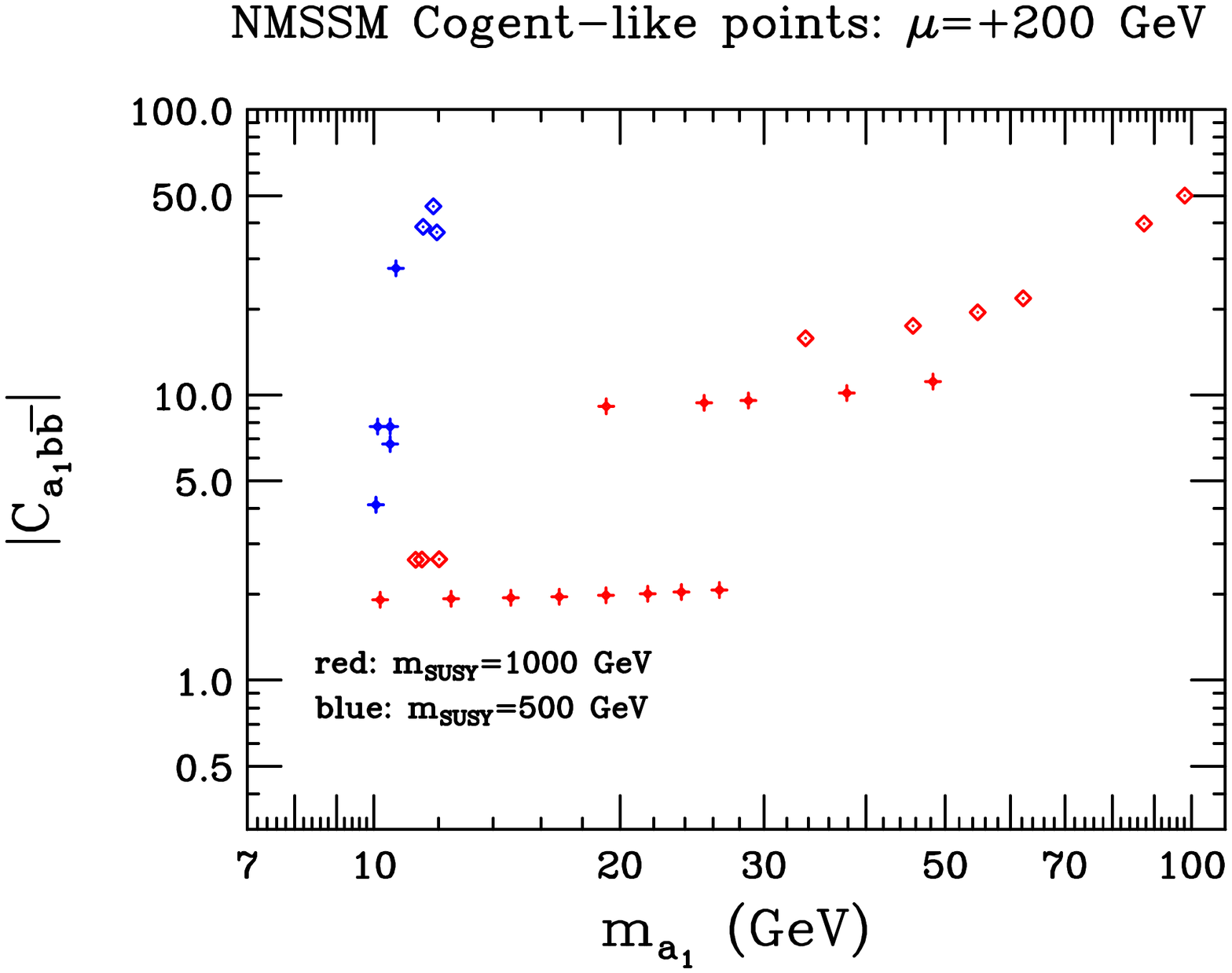}
\end{center}
\caption{Left plot: $\mai$ vs. $\mcnone$ for $\mueff=+200\gev$ points
  satisfying level-I constraints. Right plot: $|\caibb|$ vs. $\mai$
  for $\mueff=+200\gev$ points satisfying level-I constraints. Parameters
  not shown are fixed as stated in the text.}
\label{ma1vs}
\end{figure}

Let us focus on a few more details regarding the $\mueff=+200\gev$
points.  As already noted, only these are fully consistent with
$(g-2)_\mu$ constraints.  As described above, they have only a small
violation of nominal $b\to s\gam$ bounds. In the left-hand plot of
Fig.~\ref{ma1vs}, we show the range of $\mai$ values as a function of
$\mcnone$. One observes the expected trend of increasing $\mai$ with
increasing $\mcnone$ needed in order to achieve appropriate relic
abundance.

As discussed in Ref.~\cite{Dermisek:2009fd}, scenarios with a light
$\ai$ can potentially be probed by directly searching for the $\ai$ at
hadron colliders.  The discovery potential is basically a function of
the strength of the $\ai b\anti b$ reduced coupling, $\caibb$. In the
NMSSM context, $\caibb=\cos\theta_A\tanb$, where $\cos\theta_A$
specifies the amount of the $\ai$ that resides in the MSSM-like
doublet sector as opposed to the singlet component:
\beq
\ai=\cos\theta_A a_{MSSM}+\sin\theta_A a_{S}
\eeq
In the absence of $\cos\theta_A$ suppression, the $\ai$ would be
strongly coupled to down-type quarks proportionally to $\tanb$.
However, many of the points with large $\sigsi$ have $\cta$ values
significantly below unity.  The right-hand plot of Fig.~\ref{ma1vs}
shows $|\caibb|$ vs. $\mai$ for all the $\mueff=+200\gev$ points.  We
see significant variation of $|\caibb|$, but find many points with
fairly large values.  Of course, the larger $|\caibb|$ is, the easier
it will be to detect the $\ai$ directly in hadronic collisions, for
example via $b\anti b\ai$ production followed by $\ai\to \tauptaum$ or
$gg\to \ai \to \mupmum$. Some of the $|\caibb|$ values are
sufficiently large that early detection at the LHC might be feasible.

Another interesting question is how well the points plotted
agree with precision electroweak constraints.  This can be assessed by
computing the effective precision electroweak mass defined by 
\beq 
\ln\meff =\sum_{i=1,2,3} |C_V(h_i)|^2 \ln m_{h_i}\,.  
\eeq 
One finds that all $\mueff=+200\gev$ points with $\msusy=1\tev$ have
$\meff\in[114\gev,116\gev]$.  In comparison, the $\msusy=500\gev$
points can have $\meff$ as low as $100\gev$, thereby achieving
excellent agreement with precision electroweak measurements. Such
points are closely related to the ``ideal'' Higgs scenarios, but are
more complex in nature.  However, as shown in Fig.~\ref{mu+200allpts},
the largest $\sigsi$ that can be achieved for such points is of order
$0.1\times 10^{-4}\pb$, a factor of $\gsim 15$ below that needed to most
naturally describe the \cogent/DAMA observations.

Although we have not explicitly performed the necessary computations,
we anticipate that the $\msusy=1000\gev$ points will have significant
electroweak symmetry breaking (EWSB) finetuning (\ie\ to predict the
correct value of $\mz$ will require very precise adjustment of the
GUT-scale soft-SUSY-breaking parameters) whereas much less finetuning
should be required in the case of the $\msusy=500\gev$ points.

\begin{table}[t!]
  \caption{Properties of a particularly attractive but
    phenomenologically complex NMSSM point with
    $\mu=+200\gev$, $\tanb=40$ and $\msusy=500\gev$.  This point
    predicts values for $\br(t\to \hp b)\times
    \br(\hp\to\tau^+\nu_\tau)$ and for $b\anti b +Higgs$ 
    production with $Higgs\to \tauptaum$ 
    (for all neutral Higgs bosons) below 
    current observed Tevatron limits. In the last row, the brackets
    give the range of predictions for this point after including
    theoretical errors as employed in NMHDECAY. \label{pt10p}}
  \smallskip
\begin{tabular}{|c|c|c|c|c|c|c|c|}
\hline
$\lam$ & $\kap$ & $\alam$ & $\akap$ & $M_1$ & $M_2$ & $M_3$ & $\asoft$
\cr
\hline
$0.081$ &   $0.01605$ & $  -36 \gev$ & $-3.25\gev $ & $8\gev$ &
$200\gev$ & $300\gev$ & $479\gev$ \cr
\hline
\end{tabular}

\begin{tabular}{|c|c|c|c|c|c|}
\hline
$\mhi$ & $\mhii$ & $\mhiii$ & $\mai$ & $\maii$ & $\mhp$ \cr
\hline
$53.8\gev$ & $97.3\gev$ & $126.2\gev$ & $10.5\gev$ & $98.9\gev$ &
$128.4\gev$  \cr 
\hline
\end{tabular}

\begin{tabular}{|c|c|c|c|}
\hline
 $C_V(\hi)$ & $C_V(\hii)$ & $C_V(\hiii)$ & $\meff$ \cr
\hline
$-0.505$ & $0.137$ & $0.852$ & $101\gev$ \cr
\hline
\end{tabular}
\begin{tabular}{|c|c|c|c|c|}
\hline
$\chibb$ & $\chiibb$ & $\chiiibb$ & $\caibb$ & $\caiibb$ \cr
\hline
$0.24$ & $39.7$ & $-5.1$ & $6.7$ & $39.4$ \cr
\hline
\end{tabular}

\begin{tabular}{|c|c|c|c|c|c|c|c|}
\hline
$\mcnone$ & $N_{11}$ & $N_{13}$ & $\mcntwo$ & $\mcpmone$ & $\sigsi$ & $\sigsd$ & $\Omega h^2$  \cr 
\hline
$7\gev$ & $-0.976$ & $-0.212$ & $79.1\gev$  & $153\gev$ & $0.93 \times 10^{-5} \pb$ & $0.45\times 10^{-4}\pb$ & $0.12$ \cr
\hline
\end{tabular}

\begin{tabular}{|c|c|c|c|c|c|c|}
\hline  
$\br(\hi\to\ai\ai)$ & $\br(\hii\to\ai\ai)$ & $\br(\hiii\to Higgs~pair)$  & $\br(\ai\to
  jj)$ & $\br(\ai\to \tauptaum)$ & $\br(\ai\to \mupmum)$ &
  $\br(\aii\to\mupmum)$ \cr 
  \hline
   $0.96$ & $0.31\times  10^{-5}$ & $0.3$  & $0.28$ & $0.79$ & $0.003$ &  $4.3\times 10^{-4}$ \cr
  \hline
\end{tabular}

\begin{tabular}{|c|c|c|c|}
\hline
$\br(B_s\to\mupmum)$ & $\br(b\to s\gam)$ & $\br(\hp\to \tau^+\nu_\tau)$
& $(g-2)_\mu$ \cr
\hline
$[1.7-6.0]\times 10^{-9}$ & $[5.8-12.5]\times  10^{-4}$ & $[0.91-4.22]\times 10^{-4}$ &
$[4.42-5.53]\times 10^{-9}$ \cr 
\hline
\end{tabular}

\caption{The $\pm 2\sigma$ experimental ranges for the $B$ physics
  observables tabulated in the last row of
  Table~\ref{pt10p}. \label{explimits}}
\begin{tabular}{|c|c|c|c|}
\hline
$\br(B_s\to\mupmum)$ & $\br(b\to s\gam)$ & $\br(\hp\to \tau^+\nu_\tau)$
& $(g-2)_\mu$ \cr
\hline
$<5.8\times 10^{-8}$ (95\% CL) & $[3.03-4.01]\times  10^{-4}$ & $[0.34-2.3]\times 10^{-4}$ &
$[0.88-4.6]\times 10^{-9}$ \cr 
\hline
\end{tabular}
\end{table}

It is perhaps interesting to give details for the $\tanb=40$,
$\msusy=500\gev$ ``semi-ideal-Higgs'' point with $\mcnone$ in the
center of the \cogent\ mass region and $\sigsi\sim 0.1\times
10^{-4}\pb$ found in Fig.~\ref{mu+200allpts}. The relevant details are
presented in Table~\ref{pt10p}. For this point it is the $\hii$ with
$\mhii\sim 97\gev$ that is mainly responsible for a substantial size
for $\sigsi$ (since $\chiibb$ is large --- see the 3rd row of
Table~\ref{pt10p}).  In contrast, the $\hi$ has relatively small
down-type quark coupling as can be seen from the tabulated value of $\chibb$.
Note that low $\meff$ is achieved despite the fact that the Higgs,
namely the $\hiii$, that carries the bulk ($74\%$) of the $WW,ZZ$
coupling-squared has mass $\mhiii\sim 126\gev$.  This is because the
$\hi$ carries about $25\%$ of the $WW,ZZ$ coupling-squared and has
very low mass.

According to the NMSSMTools package, the only statistically
significant Higgs signal for this point in the normal LHC search
channels arises in the $WW\to\hiii\to \tauptaum$ channel where one
finds statistical significances relative to background of $3.8$ and
$14$ at low and high luminosity, respectively.  Even though the
$\hiii$ in this scenario is fairly SM-like ($|C_V(\hiii)|^2\sim 0.72$)
its decays to $WW,ZZ$ and $\gam\gam$ are suppressed to levels well
below those typical of the SM Higgs of the same (low) mass, partly because
of the smaller $|C_V(\hiii)|^2$ and partly because of significant
$\hiii\to Higgs~pair$ decays. In addition to the
$WW\to\hiii\to\tauptaum$ LHC signal, it seems to us that the $b\anti b
\hii(\to\tauptaum)$ signal would also be strong.  One
would also wish to push discovery of the $\ai$ in the $gg\to \ai\to
\mupmum$ channel --- the preliminary estimates of
\cite{Dermisek:2009fd} indicate this signal might well be observable
given the relatively large value of $\caibb$ tabulated above, despite
the fact that $\mai$ is in the Upsilon mass region.

\begin{figure}[h!]
\begin{center}
\hspace*{-.3in}\includegraphics[width=0.5\textwidth,angle=90]{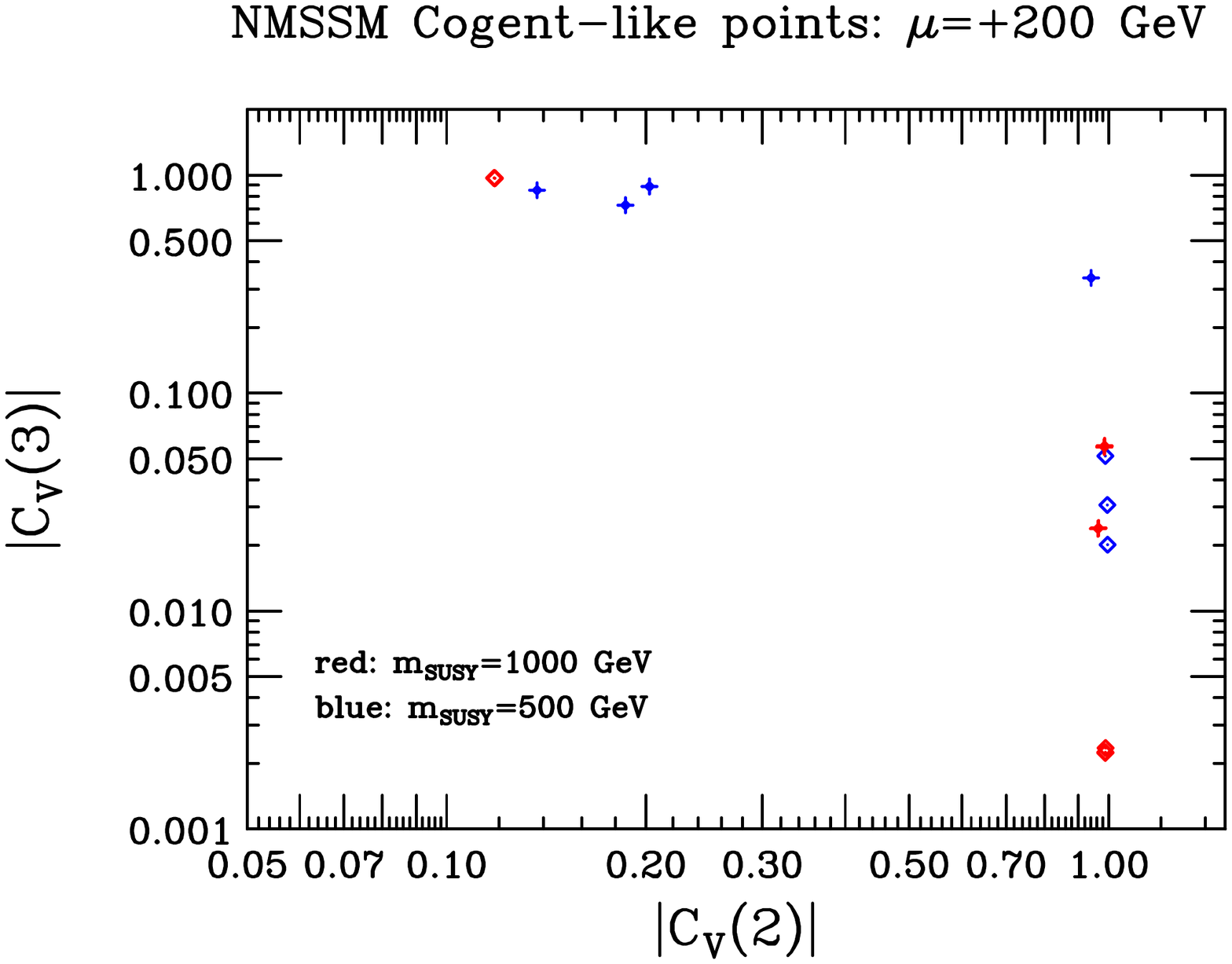}\hspace*{-1in}\includegraphics[width=0.5\textwidth,angle=90]{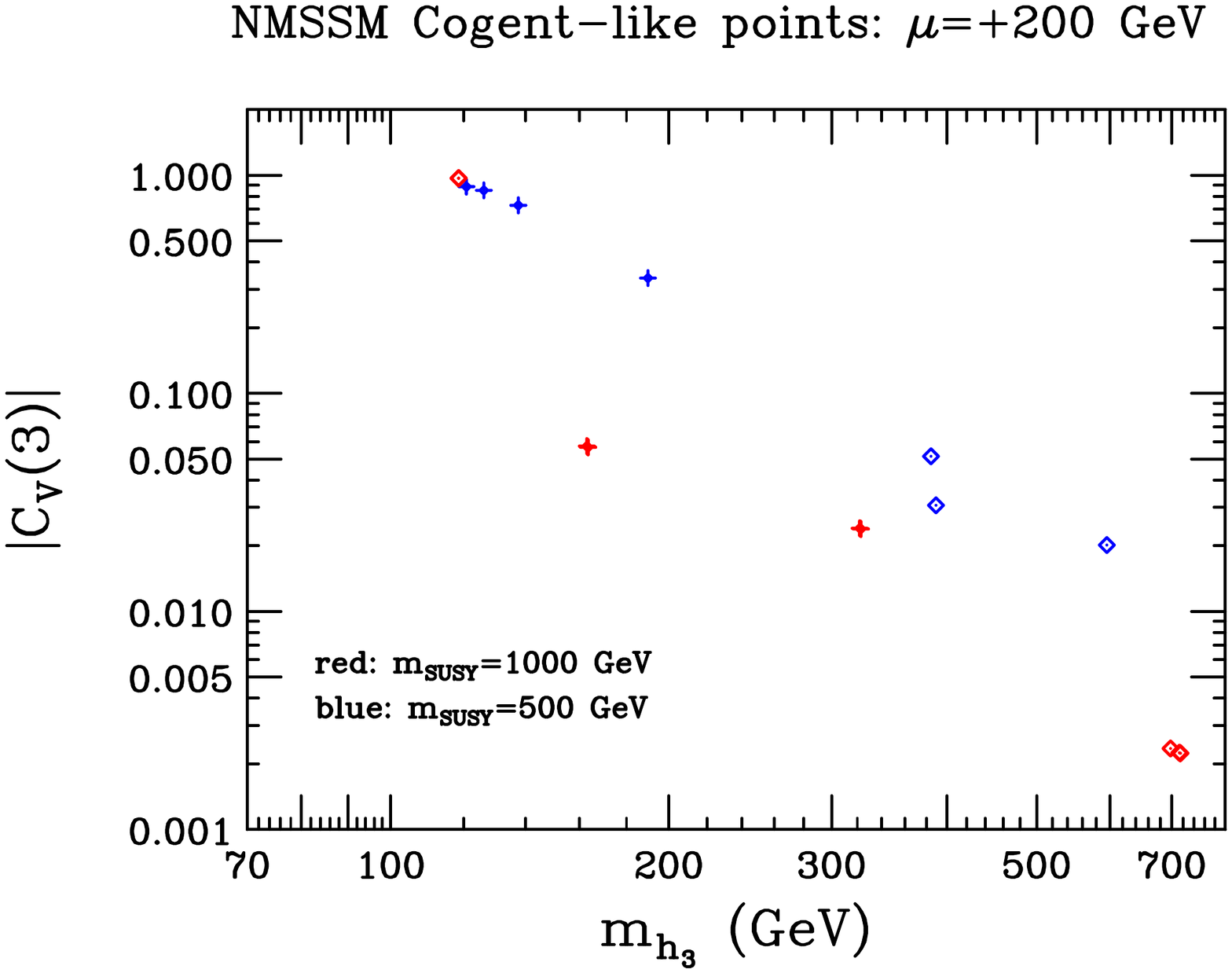}
\end{center}
\caption{$|C_V(3)|$ vs. $|C_V(2)|$ and $\mhiii$ for $\mueff=+200$ points.
 Parameters not shown
  are fixed as stated in the text. Only level-I constraints are
  imposed. There is a great amount of point overlap in this plot.}
\label{mh3plotsmu+200}
\end{figure}

This and other similar points for which $\hiii$ is the SM-like Higgs
appear distinctly in Fig.~\ref{mh3plotsmu+200}. In these scenarios,
LEP constraints are easily evaded for the $\hi$ and $\hii$ since they
have greatly reduced $WW,ZZ$ coupling, and in the case of the $\hi$
the dominance of $\hi\to \ai\ai\to 4\tau$ decays greatly reduces LEP
sensitivity as well. LEP constraints on the $\hiii$ do not enter since
$\mhiii>114\gev$ for these cases.

As regards the $B$ physics results in the last row of
Table~\ref{pt10p}, the possible range of predictions is that obtained
by taking the central prediction of the point after subtracting or
adding the theoretical error.  These ranges can be compared to the
current $\pm 2\sigma$ experimental ranges of Table~\ref{explimits}.
For all but $\br(b\to s\gam)$ there is satisfactory overlap of the
predicted range with the experimental range. If we quantify the
discrepancy between the predicted and observed ranges as described
earlier, the overlap failure is at about the $0.5\sigma$ level.

\begin{figure}[h!]
\begin{center}
\includegraphics[width=0.65\textwidth,angle=90]{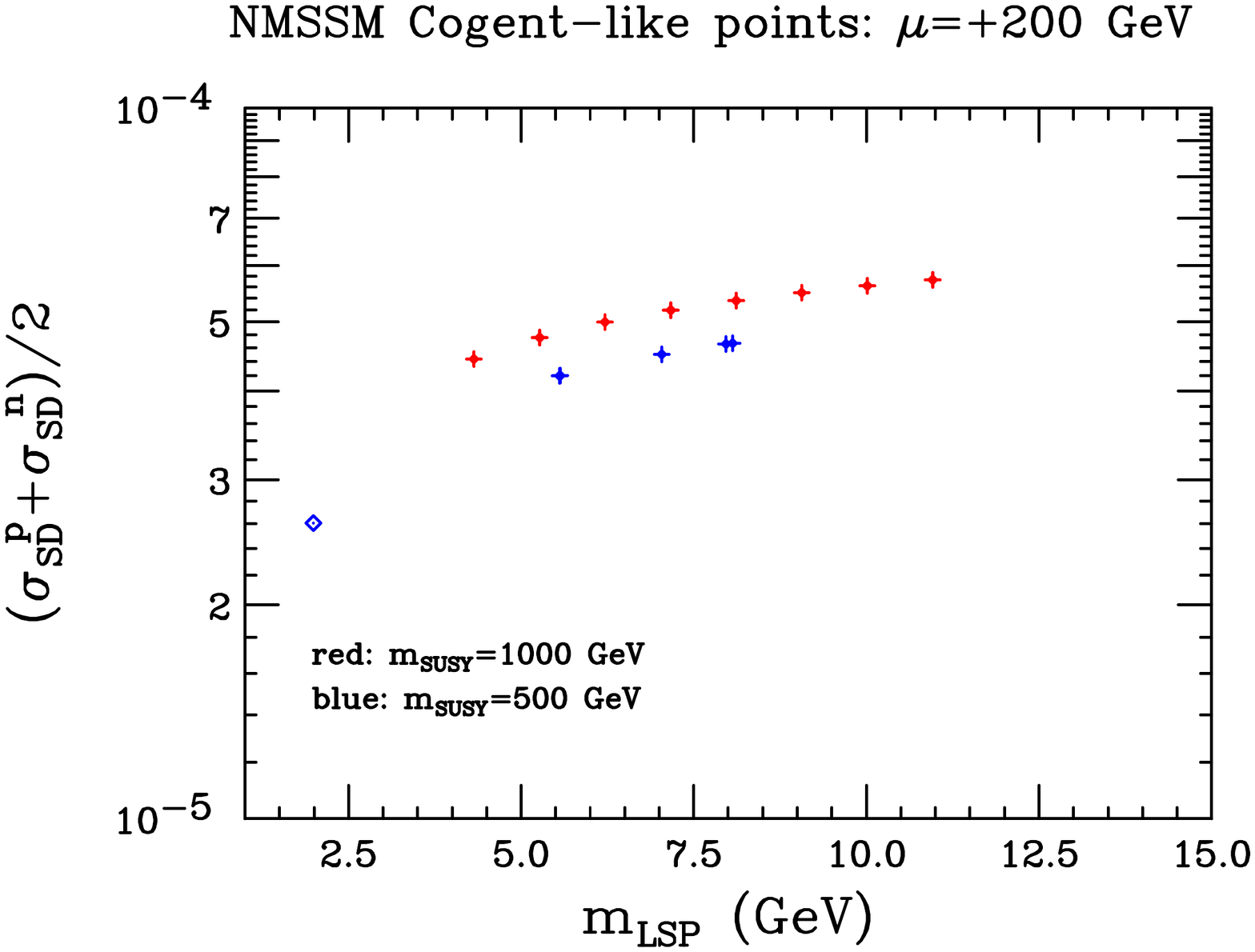}
\end{center}
\caption{$\sigsd$ vs. $\mcnone$ for all $\mueff=+200\gev$ points
  satisfying level-I constraints.}
\label{sigsdvsmlsp}
\end{figure}

Let us briefly discuss the spin-dependent cross sections for the
$\mu=+200\gev$ points. These are basically only a function of $\tanb$
and $\mcnone$. The proton and neutron spin-dependent cross sections
are very similar in magnitude.  Thus, we confine ourselves to plotting
the average value $\sigsd\equiv (\sig_{SD}^p+\sig_{SD}^n)/2$ in
Fig.~\ref{sigsdvsmlsp}, even though it is only the separate cross
sections that are directly experimentally measurable.  One finds that
$\sigsd$ varies from a low near $0.24\times 10^{-4}\pb$ for
$\mcnone\sim2.5\gev$ to a high of $\sim 0.6\times 10^{-4}$ for
$\mcnone\sim 11\gev$ (the largest value we have considered for
$\mueff=+200\gev$ points).

All the cross section results obtained above are based on the nominal
NMSSMTools and micrOMEGAs assumptions.  It is worth mentioning several
means of enhancing these cross sections.  First, we note that the
cross section magnitudes have assumed the standard $s$-quark content
for the proton.  In \cite{Das:2010ww}, the possibility of enhancing
$\sigsi$ by increasing the $s$-quark content of the nucleon was
discussed. In particular, if one changes the nominal micrOMEGAs values
of $\sigma_{\pi N}=55\mev$, $\sig_0=35\mev$ to $\sigma_{\pi
  N}=73\mev$, $\sig_0=30\mev$ then $\sigsi$ will be enhanced by
roughly a factor of 3.3.  We believe that such a large shift is not
consistent with current constraints and lattice calculations.  At
most, one might consider $\sigma_{\pi N}\sim 60\mev$ and
$\sigma_0=30\mev$~\cite{leutwyler}, leading to an enhancement of about
50\%.  In fact, the preponderance of information suggests that, if
anything, a lower value of $\sigma_{\pi N}\sim 50\mev$ is preferred
leading to a decrease in the nucleon's $s$-quark content and thereby a
decrease in $\sigsi$. Another possibility is to employ the larger
average local dark matter density $\rho=[0.4-0.485]$~GeV/cm$^3$
suggested in recent papers (see the summary of \cite{Pato:2010yq})
instead of the micrOMEGAs default value of $0.3$~GeV/cm$^3$. This
would result in a $\sim 60\%$ decrease in the $\sigsi$ required to
explain the \cogent/DAMA events. Using both a 50\% $s$-quark
enhancement and the larger $\rho$ one could get about a factor of 2
decrease in the discrepancy between the NMSSM 
predictions for $\sigsi$ and the $\sigsi$ values needed to describe the
\cogent/DAMA observations. 

For nominal $s$-quark content, our results differ
somewhat from the NMSSM scan performed in~\cite{Das:2010ww}. Their
results for $\sigsi$ for $\mueff>0$ are roughly a factor of 10 below
ours. We believe that this is primarily because in their scenarios the
$\hi$ is always SM-like, whereas in our highest-$\sigsi$ cases the
$\hi$ has enhanced down-type quark coupling and it is the $\hii$ or
$\hiii$ that is SM-like. This means that their non-SM-like mainly
$H_d$-like Higgs, the $\hii$ in their case, is typically significantly
heavier than in our scenarios. Since $\sigsi\propto 1/m_{H_d-like}^4$
a factor of 10 increase in $\sigsi$ can be achieved if $m_{H_d-like}$
is decreased by a factor of $1.77$.  For their scans, the largest
$\sigsi$ is achieved for $m_{H_d-like}\in[205\gev,260\gev]$, whereas
our large $\sigsi$ values typically have $m_{H_d-like}\leq
100\gev$. They obtain some gain in cross section since their typical
$\mueff$ is lower ($\sim 138\gev$ vs. our $200\gev$), leading to
somewhat larger $N_{13}^2$.  The larger $m_{H_d-like}$ in the scans of
\cite{Das:2010ww} imply a larger $\mhp$ with the consequence that
their largest $\sigsi$ points are within the nominal $\pm 2\sigma$
constraints from $b\to s\gam$ whereas our high-$\sigsi$, $\mueff>0$
points are about $0.5\sigma$ outside the $\pm 2\sigma$ region.

\section{Summary and Conclusions}
\label{conclusions}

We have examined parameter choices within the NMSSM that are
potentially capable of yielding a large spin-independent cross section
for nucleon-LSP scattering at low LSP mass, consistent with that
needed to describe the \cogent/DAMA observations. We have required
that all LEP and BaBar constraints be satisfied and that accepted
points have correct relic density and sufficiently small
$\br(B_s\to\mupmum)$.  We have then examined the impact of additional
constraints associated with Tevatron observations, other $B$ physics
observations and $(g-2)_\mu$.

For standard assumptions regarding the $s$-quark content of the
nucleons, we have found that in the NMSSM the largest spin-independent
cross section that can be achieved for a relevant range of $\mcnone$
if $\mueff>0$ is roughly a factor of 10 to 20 shy of that needed to
describe the \cogent/DAMA event excesses assuming standard relic
density, the latter corresponding to $\sigsi\sim (1.4-3.5)\times
10^{-4}\pb$. In particular, $\sigsi$ for $\mueff=+200\gev$ can be no
larger than $0.14\times 10^{-4}\pb$ after imposing the Tevatron
constraints (but allowing for a very mild violation in $b\to s\gam$).
If one allows for $\rho \sim [0.4-0.485]$~GeV/cm$^3$ instead of
$0.3$~GeV/cm$^3$ this will decrease the $\sigsi$ required to explain
\cogent/DAMA by about 60\% to perhaps as low as $\sim 10^{-4}\pb$.
Nonetheless, our maximal $\sigsi$ values, of order $0.14\times
10^{-4}$ for $\mueff=+200\gev$, would still be well shy of that
needed. There is also some uncertainty in the $s$-quark nucleon
content. It is possible to suppose that it could be enhanced by about
50\%, although a 50\% decrease is perhaps even more
reasonable. Combining a 50\% increase with the larger $\rho$, one
would still be a factor of at least 5 short of explaining the
\cogent/DAMA event rates.

For standard $s$-quark nucleon content, the largest $\sigsi$ values
found for $\mueff<0$ are $\sim 0.6\times 10^{-4}\pb$, within a factor
of 3 to 5 of the needed (assuming nominal $\rho=0.3$~GeV/cm$^3$)
$\sigsi =(1.4-3.5)\times 10^{-4}\pb$.  Unfortunately, $\mueff<0$ NMSSM
parameter choices yielding such large $\sigsi$ all predict an
anomalous magnetic moment for the muon that is strongly discrepant
with the observed $(g-2)_\mu$.  Nonetheless, it is not impossible that
there is some resolution of this disagreement coming from physics
beyond the NMSSM.

We have illustrated that Tevatron (and, presumably soon, the LHC)
constraints on $b\anti b+Higgs$ production and $t\to \hp b$ decays are
highly relevant for constraining large-$\sigsi$ scenarios.  Thus, it
is clear that if the \cogent\ observations really are dark matter
detection and if the NMSSM is the relevant model, detection of one or
more of the $\ai$, $\hi$, $\aii$ and $\hp$ of the NMSSM at the
Tevatron and LHC should be close at hand in the above channels.
However, it is also the case that detecting the SM-like Higgs of these
scenarios will be very difficult.

On another front, in a companion paper~\cite{Belikov:2010yi} we have
demonstrated that allowing an extension of the NMSSM to include
additional superpotential terms and/or soft-SUSY-breaking terms (while
still keeping just one singlet superfield) will be sufficiently less
constraining that $\sigsi$ values large enough to describe the
\cogent\ excess can be achieved without {\it any} LEP, Tevatron,
BaBar, $B$-physics (other than a quite small $b\to s\gam$ deviation)
or $(g-2)_\mu$ issues, and using nominal $s$-quark nucleon content and
standard relic density $\rho$. The key new feature is that the
additional parameters allow scenarios consistent with all constraints
for which the $\cnone$ is highly singlet and the $\hi$ is largely
singlet-like with large $\cnone\cnone\hi$ coupling and low $\mhi$.

\section*{Acknowledgements} We would like to thank S. Chang for
several helpful conversations.  DH is supported
by the US Department of Energy, including grant DE-FG02-95ER40896, and
by NASA grant NAG5-10842.  JFG is supported by US DOE grant
DE-FG03-91ER40674.  JFG and DH also received support from the Aspen Center for
Physics while working on this project.


\begin{thebibliography}{9}


\bibitem{cogentnew}
C.~E.~Aalseth {\it et al.} [The CoGeNT Collaboration],
arXiv:1002.4703 [astro-ph.CO].

\bibitem{liam}
  A.~L.~Fitzpatrick, D.~Hooper and K.~M.~Zurek,
  %``Implications of CoGeNT and DAMA for Light WIMP Dark Matter,''
  arXiv:1003.0014 [hep-ph].
  %%CITATION = ARXIV:1003.0014;%%

\bibitem{Kopp:2009qt}
  J.~Kopp, T.~Schwetz and J.~Zupan,
  %``Global interpretation of direct Dark Matter searches after CDMS-II
  %results,''
  JCAP {\bf 1002}, 014 (2010)
  [arXiv:0912.4264 [hep-ph]].
  %%CITATION = JCAPA,1002,014;%%



\bibitem{Chang:2010yk}
  S.~Chang, J.~Liu, A.~Pierce, N.~Weiner and I.~Yavin,
  %``CoGeNT Interpretations,''
  arXiv:1004.0697 [hep-ph].
  %%CITATION = ARXIV:1004.0697;%%



    %\cite{Bernabei:2010mq}
\bibitem{DAMAnew}
  R.~Bernabei {\it et al.},
  %``New results from DAMA/LIBRA,''
  arXiv:1002.1028 [astro-ph.GA].
  %%CITATION = ARXIV:1002.1028

%\cite{Hooper:2010uy}
\bibitem{Hooper:2010uy}
  D.~Hooper, J.~I.~Collar, J.~Hall and D.~McKinsey,
  %``A Consistent Dark Matter Interpretation For CoGeNT and DAMA/LIBRA,''
  arXiv:1007.1005 [hep-ph].
  %%CITATION = ARXIV:1007.1005;%%

%\cite{Sorensen:2010hq}
%\bibitem{Sorensen:2010hq}
%  P.~Sorensen,
  %``A coherent understanding of low-energy nuclear recoils in liquid xenon,''
%  arXiv:1007.3549 [astro-ph.IM].
  %%CITATION = ARXIV:1007.3549;%%




\bibitem{Kuflik:2010ah}
  E.~Kuflik, A.~Pierce and K.~M.~Zurek,
  %``Light WIMPs: the Largest Detection Scattering Cross Sections in the MSSM,''
  arXiv:1003.0682 [hep-ph].
  %%CITATION = ARXIV:1003.0682;%%

\bibitem{Feldman:2010ke}
  D.~Feldman, Z.~Liu and P.~Nath,
  %``Low Mass Neutralino Dark Matter in the MSSM with Constraints from $B_s\to
  %\mu^+\mu^-$ and Higgs Search Limits,''
  arXiv:1003.0437 [hep-ph].
  %%CITATION = ARXIV:1003.0437;%%



\bibitem{lightLSP}
  D.~Hooper and T.~Plehn,
%  ``Supersymmetric dark matter: How light can the LSP be?,''
  Phys.\ Lett.\  B {\bf 562}, 18 (2003)
  [arXiv:hep-ph/0212226];
  %%CITATION = PHLTA,B562,18;%%
  A.~Bottino, N.~Fornengo and S.~Scopel,
  %``Light relic neutralinos,''
Phys.\ Rev.\  D {\bf 67}, 063519 (2003)
[arXiv:hep-ph/0212379].
%%CITATION = PHRVA,D67,063519;%%

%\cite{Belikov:2010yi}
\bibitem{Belikov:2010yi}
  A.~V.~Belikov, J.~F.~Gunion, D.~Hooper and T.~M.~P.~Tait,
  %``CoGeNT, DAMA, and Light Neutralino Dark Matter,''
  arXiv:1009.0549 [hep-ph].
  %%CITATION = ARXIV:1009.0549;%%

%\cite{Pato:2010yq}
\bibitem{Pato:2010yq}
  M.~Pato, O.~Agertz, G.~Bertone, B.~Moore and R.~Teyssier,
  %``Systematic uncertainties in the determination of the local dark matter
  %density,''
  Phys.\ Rev.\  D {\bf 82}, 023531 (2010)
  [arXiv:1006.1322 [astro-ph.HE]].
  %%CITATION = PHRVA,D82,023531;%%








\bibitem{wmap}
  E.~Komatsu {\it et al.}  [WMAP Collaboration],
  %``Five-Year Wilkinson Microwave Anisotropy Probe (WMAP\altaffilmark 1 )
  %Observations:Cosmological Interpretation,''
  Astrophys.\ J.\ Suppl.\  {\bf 180}, 330 (2009)
  [arXiv:0803.0547 [astro-ph]].
  %%CITATION = APJSA,180,330;%%


\bibitem{nmssm}
  J.~R.~Ellis, J.~F.~Gunion, H.~E.~Haber, L.~Roszkowski and F.~Zwirner,
  %``Higgs Bosons in a Nonminimal Supersymmetric Model,''
  Phys.\ Rev.\  D {\bf 39}, 844 (1989);
  %%CITATION = PHRVA,D39,844;%%
  H.~P.~Nilles, M.~Srednicki and D.~Wyler,
  %``Weak Interaction Breakdown Induced By Supergravity,''
  Phys.\ Lett.\  B {\bf 120}, 346 (1983);
  %%CITATION = PHLTA,B120,346;%%
  J.~E.~Kim and H.~P.~Nilles,
  %``The Mu Problem And The Strong CP Problem,''
  Phys.\ Lett.\  B {\bf 138}, 150 (1984);
  %%CITATION = PHLTA,B138,150;%%
  M.~Drees,
  %``Supersymmetric Models with Extended Higgs Sector,''
  Int.\ J.\ Mod.\ Phys.\  A {\bf 4}, 3635 (1989).
  %%CITATION = IMPAE,A4,3635;%%


%\cite{Dermisek:2005ar}
\bibitem{Dermisek:2005ar}
  R.~Dermisek and J.~F.~Gunion,
  %``Escaping the large fine tuning and little hierarchy problems in the  next
  %to minimal supersymmetric model and h --> a a decays,''
  Phys.\ Rev.\ Lett.\  {\bf 95}, 041801 (2005)
  [arXiv:hep-ph/0502105].
  %%CITATION = PRLTA,95,041801;%%
  
  %\cite{Funakubo:2002yb}
\bibitem{Funakubo:2002yb}
%\cite{Espinosa:1993bs}
%\bibitem{Espinosa:1993bs}
  J.~R.~Espinosa and M.~Quiros,
  %``The Electroweak phase transition with a singlet,''
  Phys.\ Lett.\  B {\bf 305}, 98 (1993)
  [arXiv:hep-ph/9301285];
  %%CITATION = PHLTA,B305,98;%%
  K.~Funakubo, S.~Tao and F.~Toyoda,
  %``CP violation in the Higgs sector and phase transition in the MSSM,''
  Prog.\ Theor.\ Phys.\  {\bf 109}, 415 (2003)
  [arXiv:hep-ph/0211238].
  %%CITATION = PTPKA,109,415;%%

%\cite{Dermisek:2006wr}
\bibitem{Dermisek:2006wr}
  R.~Dermisek and J.~F.~Gunion,
  %``The NMSSM close to the R-symmetry limit and naturalness in h --> aa  decays
  %for m(a) < 2m(b),''
  Phys.\ Rev.\  D {\bf 75}, 075019 (2007)
  [arXiv:hep-ph/0611142].
  %%CITATION = PHRVA,D75,075019;%%
  
 
%\cite{gunion}
\bibitem{gunion}
  J.~F.~Gunion, D.~Hooper and B.~McElrath,
  %``Light neutralino dark matter in the NMSSM,''
  Phys.\ Rev.\  D {\bf 73}, 015011 (2006)
  [arXiv:hep-ph/0509024];
  %%CITATION = PHRVA,D73,015011;%%

%\cite{Ellwanger:2004xm}
\bibitem{Ellwanger:2004xm}
  U.~Ellwanger, J.~F.~Gunion and C.~Hugonie,
  %``NMHDECAY: A Fortran code for the Higgs masses, couplings and decay  widths
  %in the NMSSM,''
  JHEP {\bf 0502}, 066 (2005)
  [arXiv:hep-ph/0406215].
  %%CITATION = JHEPA,0502,066;%%



%\cite{Ellwanger:2005dv}
\bibitem{Ellwanger:2005dv}
  U.~Ellwanger and C.~Hugonie,
  %``NMHDECAY 2.0: An Updated program for sparticle masses, Higgs masses,
  %couplings and decay widths in the NMSSM,''
  Comput.\ Phys.\ Commun.\  {\bf 175}, 290 (2006)
  [arXiv:hep-ph/0508022].
  %%CITATION = CPHCB,175,290;%%

%\cite{Belanger:2006is}
\bibitem{Belanger:2006is}
  G.~Belanger, F.~Boudjema, A.~Pukhov and A.~Semenov,
  %``micrOMEGAs2.0: A program to calculate the relic density of dark matter  in
  %a generic model,''
  Comput.\ Phys.\ Commun.\  {\bf 176}, 367 (2007)
  [arXiv:hep-ph/0607059].
  %%CITATION = CPHCB,176,367;%%

%\cite{nmssmtools}
\bibitem{nmssmtools}
U. Ellwanger and C. Hugonie, 
http://www.th.u-psud.fr/NMHDECAY/nmssmtools.html.

%\cite{Das:2010ww}
\bibitem{Das:2010ww}
  D.~Das and U.~Ellwanger,
  %``Light dark matter in the NMSSM: upper bounds on direct detection cross
  %sections,''
  arXiv:1007.1151 [hep-ph].
  %%CITATION = ARXIV:1007.1151;%%


%%%%%%%

%\cite{Schael:2010aw}
\bibitem{Schael:2010aw}
  S.~Schael {\it et al.}  [ALEPH Collaboration],
  %``Search for neutral Higgs bosons decaying into four taus at LEP2,''
  JHEP {\bf 1005}, 049 (2010)
  [arXiv:1003.0705 [hep-ex]].
  %%CITATION = JHEPA,1005,049;%%

%\cite{Benjamin:2010xb}
\bibitem{Benjamin:2010xb}
  D.~Benjamin {\it et al.}  [Tevatron New Phenomena \& Higgs Working Group],
  %``Combined CDF and D0 upper limits on MSSM Higgs boson production in tau-tau
  %final states with up to 2.2 fb-1,''
  arXiv:1003.3363 [hep-ex].
  %%CITATION = ARXIV:1003.3363;%%

%\cite{:2009zh}
\bibitem{:2009zh}
  V.~M.~Abazov {\it et al.}  [D0 Collaboration],
  %``Search for charged Higgs bosons in top quark decays,''
  Phys.\ Lett.\  B {\bf 682}, 278 (2009)
  [arXiv:0908.1811 [hep-ex]].
  %%CITATION = PHLTA,B682,278;%%

%\cite{Dermisek:2010mg}
\bibitem{Dermisek:2010mg}
  R.~Dermisek and J.~F.~Gunion,
  %``New constraints on a light CP-odd Higgs boson and related NMSSM Ideal Higgs
  %Scenarios,''
  Phys.\ Rev.\  D {\bf 81}, 075003 (2010)
  [arXiv:1002.1971 [hep-ph]].
  %%CITATION = PHRVA,D81,075003;%%

%\cite{Dermisek:2009fd}
\bibitem{Dermisek:2009fd}
  R.~Dermisek and J.~F.~Gunion,
  %``Direct production of a light CP-odd Higgs boson at the Tevatron and LHC,''
  Phys.\ Rev.\  D {\bf 81}, 055001 (2010)
  [arXiv:0911.2460 [hep-ph]].
  %%CITATION = PHRVA,D81,055001;%%

\bibitem{leutwyler}  We wish to thank H. Leutwyler for very useful
  communications regarding these issues.

\end{thebibliography}
\end{document}

%%%%%% below not yet used

\bibitem{Ahmed:2009zw}
  Z.~Ahmed {\it et al.}  [The CDMS-II Collaboration],
  %``Results from the Final Exposure of the CDMS II Experiment,''
  arXiv:0912.3592 [astro-ph.CO].
  %%CITATION = ARXIV:0912.3592;%%

\bibitem{xenonnull}
 J.~Angle {\it et al.}  [XENON Collaboration],
  %``First Results from the XENON10 Dark Matter Experiment at the Gran Sasso
  %National Laboratory,''
  Phys.\ Rev.\ Lett.\  {\bf 100}, 021303 (2008)
  [arXiv:0706.0039 [astro-ph]].
  %%CITATION = PRLTA,100,021303;%%

    %\cite{Akerib:2005kh}

  %\cite{Petriello:2008jj}
\bibitem{petriello}
  F.~Petriello and K.~M.~Zurek,
  %``DAMA and WIMP dark matter,''
  JHEP {\bf 0809}, 047 (2008)
  [arXiv:0806.3989 [hep-ph]].
  %%CITATION = JHEPA,0809,047;%
  
  %\cite{Chang:2008xa}
\bibitem{Chang}
  S.~Chang, A.~Pierce and N.~Weiner,
  %``Using the Energy Spectrum at DAMA/LIBRA to Probe Light Dark Matter,''
  Phys.\ Rev.\  D {\bf 79}, 115011 (2009)
  [arXiv:0808.0196 [hep-ph]].
  %%CITATION = PHRVA,D79,115011;%%
  
  %\cite{Fairbairn:2008gz}
\bibitem{Fairbairn}
  M.~Fairbairn and T.~Schwetz,
  %``Spin-independent elastic WIMP scattering and the DAMA annual modulation
  %signal,''
  JCAP {\bf 0901}, 037 (2009)
  [arXiv:0808.0704 [hep-ph]].
  %%CITATION = JCAPA,0901,037;%%
  
  %\cite{Savage:2009mk}
\bibitem{savage}
  C.~Savage, K.~Freese, P.~Gondolo and D.~Spolyar,
  %``Compatibility of DAMA/LIBRA dark matter detection with other searches in
  %light of new Galactic rotation velocity measurements,''
  JCAP {\bf 0909}, 036 (2009)
  [arXiv:0901.2713 [astro-ph]].
  %%CITATION = JCAPA,0909,036;%%
  C.~Savage, G.~Gelmini, P.~Gondolo and K.~Freese,
  %``Compatibility of DAMA/LIBRA dark matter detection with other searches,''
  JCAP {\bf 0904}, 010 (2009)
  [arXiv:0808.3607 [astro-ph]].
  %%CITATION = JCAPA,0904,010;%%
  
  %\cite{Drobyshevski:2007zj}
\bibitem{channeling}
  E.~M.~Drobyshevski,
  %``Channeling Effect and Improvement of the Efficiency of Charged Particle
  %Registration with Crystal Scintillators,''
  Mod.\ Phys.\ Lett.\  A {\bf 23}, 3077 (2008)
  [arXiv:0706.3095 [physics.ins-det]].
  %%CITATION = MPLAE,A23,3077;%%

\bibitem{Fricke}
  G.~Fricke, C.~Bernhardt, K.~Heilig, L.~A.~Schaller, L.~Schellenberg, E.~B.~Shera and C.~W.~de Jager,
  %``Nuclear Ground State Charge Radii From Electromagnetic Interactions,''
  Atom.\ Data Nucl.\ Data Tabl.\  {\bf 60}, 177 (1995).
  %%CITATION = ADNDA,60,177;%%

\bibitem{Bernabei:2007hw}
  R.~Bernabei {\it et al.},
  %``Possible implications of the channeling effect in NaI(Tl) crystals,''
  Eur.\ Phys.\ J.\  C {\bf 53}, 205 (2008)
  [arXiv:0710.0288 [astro-ph]].
  %%CITATION = EPHJA,C53,205;%%

%\cite{Manzur:2009hp}
\bibitem{Leff}
  A.~Manzur, A.~Curioni, L.~Kastens, D.~N.~McKinsey, K.~Ni and T.~Wongjirad,
  %``Scintillation efficiency and ionization yield of liquid xenon for
  %mono-energetic nuclear recoils down to 4 keV,''
  arXiv:0909.1063 [physics.ins-det].
  %%CITATION = ARXIV:0909.106

 \bibitem{wimpless}
    J.~L.~Feng and J.~Kumar,
  %``The Wimpless Miracle: Dark-Matter Particles Without Weak-Scale Masses Or
  %Weak Interactions,''
  Phys.\ Rev.\ Lett.\  {\bf 101}, 231301 (2008)
  [arXiv:0803.4196 [hep-ph]].
  %%CITATION = PRLTA,101,231301;%%
    J.~L.~Feng, J.~Kumar and L.~E.~Strigari,
  %``Explaining the DAMA Signal with WIMPless Dark Matter,''
  Phys.\ Lett.\  B {\bf 670}, 37 (2008)
  [arXiv:0806.3746 [hep-ph]].
  %%CITATION = PHLTA,B670,37;%%

\bibitem{modelindependent}
  M.~Beltran, D.~Hooper, E.~W.~Kolb and Z.~C.~Krusberg,
  %``Deducing the nature of dark matter from direct and indirect detection
  %experiments in the absence of collider signatures of new physics,''
  Phys.\ Rev.\  D {\bf 80}, 043509 (2009)
  [arXiv:0808.3384 [hep-ph]].
  %%CITATION = PHRVA,D80,043509;%%

\bibitem{cdfdzero}
  V.~M.~Abazov {\it et al.}  [D0 Collaboration],
  %``Search for the associated production of a b quark and a neutral
  %supersymmetric Higgs boson which decays to tau pairs,''
  arXiv:0912.0968 [hep-ex];
  %%CITATION = ARXIV:0912.0968;%%
  V.~M.~Abazov {\it et al.}  [D0 Collaboration],
  %``Search for neutral Higgs bosons $tan\beta$ in the b(h/H/A) $\to b \tau
  %\tau$ channel ,''
  Phys.\ Rev.\ Lett.\  {\bf 102}, 051804 (2009)
  [arXiv:0811.0024 [hep-ex]];
  %%CITATION = PRLTA,102,051804;%%
  V.~M.~Abazov {\it et al.}  [D0 Collaboration],
  %``Search for neutral Higgs bosons in multi-b-jet events in $p \bar{p}$
  %collisions at $\sqrt{s}$ = 1.96-TeV,''
  Phys.\ Rev.\ Lett.\  {\bf 101}, 221802 (2008)
  [arXiv:0805.3556 [hep-ex]];
  %%CITATION = PRLTA,101,221802;%%
  V.~M.~Abazov {\it et al.}  [D0 Collaboration],
  %``Search for Higgs bosons decaying to $\tau$ pairs in $p \bar{p}$ collisions
  %with the D0 detector,''
  Phys.\ Rev.\ Lett.\  {\bf 101}, 071804 (2008)
  [arXiv:0805.2491 [hep-ex]];
  %%CITATION = PRLTA,101,071804;%%
  T.~Aaltonen {\it et al.}  [CDF Collaboration],
  %``Search for Higgs bosons predicted in two-Higgs-doublet models via decays to
  %tau lepton pairs in 1.96 TeV proton-antiproton collisions,''
  arXiv:0906.1014 [hep-ex];
  %%CITATION = ARXIV:0906.1014;%%
  D.~E.~Acosta {\it et al.}  [CDF Collaboration],
  %``A search for supersymmetric Higgs bosons in the di-tau decay mode in
  %$p\bar{p}$ collisions at $\sqrt{s} = 1.8$ TeV,''
  Phys.\ Rev.\  D {\bf 72}, 072004 (2005)
  [arXiv:hep-ex/0506042].
  %%CITATION = PHRVA,D72,072004;%%

    %\cite{Bottino:2002ry}
\bibitem{italians}
    A.~Bottino, F.~Donato, N.~Fornengo and S.~Scopel,
  %``Interpreting the recent results on direct search for dark matter particles
  %in terms of relic neutralino,''
  Phys.\ Rev.\  D {\bf 78}, 083520 (2008)
  [arXiv:0806.4099 [hep-ph]].
    V.~Niro, A.~Bottino, N.~Fornengo and S.~Scopel,
  %``Investigating light neutralinos at neutrino telescopes,''
  Phys.\ Rev.\  D {\bf 80}, 095019 (2009)
  [arXiv:0909.2348 [hep-ph]].
  %%CITATION = PHRVA,D80,095019;%%
    A.~Bottino, F.~Donato, N.~Fornengo and S.~Scopel,
  %``Do current WIMP direct measurements constrain light relic neutralinos?,''
  Phys.\ Rev.\  D {\bf 72}, 083521 (2005)
  [arXiv:hep-ph/0508270].
  %%CITATION = PHRVA,D72,083521;%%

\bibitem{ferrer}
  F.~Ferrer, L.~M.~Krauss and S.~Profumo,
  %``Indirect detection of light neutralino dark matter in the NMSSM,''
  Phys.\ Rev.\  D {\bf 74}, 115007 (2006)
  [arXiv:hep-ph/0609257].
  %%CITATION = PHRVA,D74,115007;%%

%%%%%%%%%%%%%%%%%%